\documentstyle[amssymb,psfig,eqsecnum,aps]{revtex}

\begin{document}
\preprint{KSU-HEP-99-001,FERMILAB-Pub-99/269-E}
\title{Evidence for Diffractive Charm Production in $\nu _{\mu }Fe$ and $\bar{\nu}
_{\mu }Fe$ Scattering at the Tevatron}
\author{T.~Adams, A.~Alton, T.~Bolton, J.~Goldman, M.~Goncharov, 
D.~Naples\footnote[1]{Present Address: University of Pittsburgh, Pittsburgh, PA, 15260}}
\address{Kansas State University, Manhattan, KS, 66506}
\author{R.~A.~Johnson, M.~Vakili,\footnote[2]{Present Address: Texas A\&M University, College Station, TX, 77843} V.~Wu}
\address{University of Cincinnati, Cincinnati, OH, 45221}
\author{J.~Conrad, J.~Formaggio, S.~Koutsoliotas,\footnote[3]{Present Address:
Bucknell University, Lewisburg, PA, 17837} J.~H.~Kim,\footnote[4]{Present 
Address: University of California, Irvine, CA, 92697} C.~McNulty, 
A.~Romosan,\footnote[5]{Present Address: University of California, Berkeley, 
CA, 94720} M.~H.~Shaevitz, P.~Spentzouris,\footnote[6]{Present Address: Fermi 
National Laboratory, Batavia, IL, 60510} E.~G.~Stern, B.~Tamminga, 
A.~Vaitaitis, E.~D.~Zimmerman}
\address{Columbia University, New York, NY, 10027}
\author{R.~H.~Bernstein, L.~Bugel, M.~J.~Lamm, W.~Marsh, 
P.~Nienaber,\footnote[7]{Present Address: Marquette University, Milwaukee, 
WI, 53201} J.~Yu}
\address{Fermi National Accelerator Laboratory, Batavia, IL, 60510}
\author{L.~de~Barbaro, D.~Buchholz, H.~Schellman, G.~P.~Zeller}
\address{Northwestern University, Evanston, IL, 60208}
\author{J.~Brau, R.~B.~Drucker, R.~Frey, D.~Mason}
\address{University of Oregon, Eugene, OR, 97403}
\author{S.~Avvakumov, P.~de~Barbaro, A.~Bodek, H.~Budd, 
D.~A.~Harris,\footnotemark[6]
K.~S.~McFarland, W.~K.~Sakumoto, U.~K.~Yang}
\address{University of Rochester, Rochester, NY, 14627}
\date{\today}
\maketitle

\begin{abstract}
We present evidence for the diffractive processes $\nu _{\mu }Fe\rightarrow
\mu ^{-}D_{S}^{+}\left( D_S^\ast \right) Fe$ and $\bar{\nu}_{\mu
}Fe\rightarrow \mu ^{+}D_{S}^{-}\left( D_S^\ast \right) Fe$ using the Fermilab
SSQT neutrino beam and the Lab E neutrino detector. We observe the neutrino
trident reactions $\nu _{\mu }Fe\rightarrow \nu _{\mu }\mu ^{-}\mu ^{+}Fe $
and $\bar{\nu}_{\mu }Fe\rightarrow \bar{\nu}_{\mu }\mu ^{+}\mu ^{-}Fe$ at
rates consistent with Standard Model expectations. We see no evidence for
neutral-current production of $J/\psi $ via either diffractive or deep
inelastic scattering mechanisms.
\vspace{0.2cm}

\centerline{PACS 13.15+g, 12.15.-y, 13.60.Le, 12.40.Vv}
\vspace{0.2cm}

\centerline{(Submitted to Physical Review D)}
\end{abstract}

\pacs{13.15+g, 12.15.-y, 13.60.Le, 12.40.Vv
{\tt$\backslash$\string
pacs\{
\}}
should
always
be
input,
even if empty.}

\draft

\addtocounter{section}{1}

\noindent\parbox{17.6cm}{
\parbox{8.6cm}{
\vspace{0.45cm}

\centerline{\bf I. INTRODUCTION}
\vspace{0.55cm}

Opposite-signed two-muon production in deep inelastic scattering (DIS) with
neutrinos or anti-neutrinos serves as a reliable signal for charm quark
production through the sequence 
\begin{equation}
\nu _{\mu }N\rightarrow \mu ^{-}DX,\ \ \text{ }D\rightarrow \mu ^{+}\nu _{\mu
}X^{\prime },
\end{equation}
where $D$ represents a stable charmed hadron. This is 
}
\hfill
\parbox{8.6cm}{
especially true in dense
targets such as the NuTeV detector 
at Fermilab where absorption of pions
and kaons 
in the hadronic shower suppresses backgrounds from meson decay
that occur via 
\begin{equation}
\nu _{\mu }N\rightarrow \mu ^{-}\pi ^{+}\left( K^{+}\right) X;\text{ }\pi
^{+}\left( K^{+}\right) \rightarrow \mu ^{+}\nu _{\mu }\text{.}
\end{equation}
Previous studies of DIS two-muon production in neutrino interactions have
yielded important measurements of the CKM matrix elements $V_{cd}$ and 
$V_{cs}$, the effective charm quark mass, and the size and shape of the
nucleon strange sea. These studies were performed in the context 
}
}

\twocolumn
\narrowtext

\noindent
\begin{figure}[tbp]                                                             
 \centerline{\psfig{figure=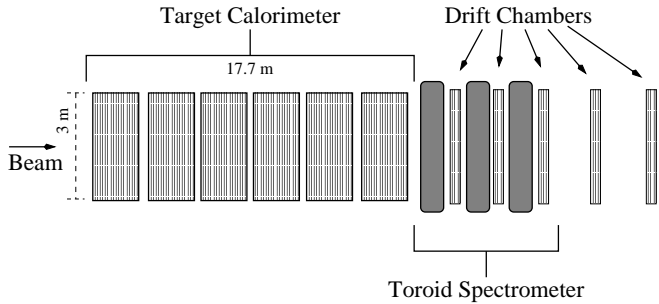,width=\columnwidth}}                        
  \vspace{0.5cm}                                                                
\caption{Schematic drawing of the Lab E detector. Beam enters from the left.    
The target/calorimeter is on the left and the toroid spectrometer is on the     
right.}                                                                         
\label{fig:labe}                                                                
\end{figure}                                                                    
                                                                                
\noindent
of the QCD-corrected quark-parton model and 
hence were restricted to large values
of energy transfer, $\nu$, to the struck nucleon resulting in large amounts
of observed hadronic energy, $E_{HAD}$.

At low hadronic energy, one expects to observe charm production via another
mechanism, namely, diffractive production of pseudoscalar and vector $D_{S}$
mesons via both coherent and incoherent scattering: 
\begin{equation}
\nu _{\mu }A\rightarrow \mu ^{-}D_{S}^{+} A, \ \
\nu _{\mu }A\rightarrow \mu ^{-}D_{S}^{+} \gamma _{\text{SOFT}}
A,
\end{equation}
where the low energy decay photon $\gamma _{\text{SOFT}}$ will accompany
production of vector $D_{S}^{\ast }$, $A=Fe$ for coherent production in our
experiment, and $A=$ $n$ or $p$ for incoherent production. Here, large
energy transfer to the meson may not result in large hadronic energy
because the nucleus remains intact. 

These reactions
have been observed at the few event level in bubble chamber and emulsion
experiments~\cite{bib:asratyan,bib:chorus}. It is important to understand 
the size of this diffractive
contribution because of its influence as a background to DIS charm
production. These processes are also of interest to future high statistics
neutrino experiments at a muon collider. They provide the possibility of
measuring the ratio $V_{cd}/V_{cs}$ via comparison of the rates $\nu _{\mu
}A\rightarrow \mu ^{-}D^{\ast +}A$ and $\nu _{\mu }A\rightarrow \mu
^{-}D_{S}^{\ast +}A$.  Additionally, they can create a signature which
mimics quasi-elastic production of $\tau -$leptons through the chain $\nu
_{\mu }A\rightarrow \mu ^{-}D_{S}^{+}\left( \gamma _{\text{SOFT}}\right) A$, 
$D_{S}^{+}\rightarrow \tau ^{+}\nu _{\tau }$ , which may be of concern to
high sensitivity $\nu _{\mu }\rightarrow \nu _{\tau }$ oscillation searches.

A competing reaction that produces the same experimental signature, an
opposite-signed muon pair with vanishing $E_{HAD}$, is the neutrino trident
process~\cite{bib:trid1,bib:trid2}: 
\begin{equation}
\nu _{\mu }A\rightarrow \mu ^{-}\mu ^{+}\nu _{\mu }A.
\end{equation}
In principle, the neutrino trident reaction provides an interesting test of
electroweak theory since contributions from $W$ and $Z$ decay produce a
reliably calculable $40\%$ destructive interference effect. In practice, the
very small cross-section implies that only a handful of neutrino tridents
have previously been observed in neutrino scattering. Furthermore, the
neutrino trident process must be considered in combination with the expected
signal from 

\begin{figure}[tbp]                                                             
\centerline{\psfig{figure=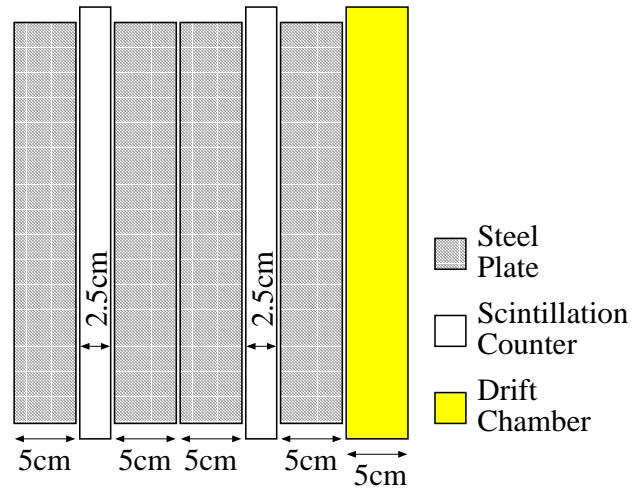,width=8.2cm}}           
\caption{Drawing of one segment of the Lab E calorimeter.}                      
\label{fig:calseg}                                                              
\end{figure} 

\noindent
diffractive charm production in experiments that are only
sensitive to two-muon final states. This point has not been recognized in
previous measurements of neutrino tridents.

NuTeV (Fermilab Experiment E815) is a high-statistics experiment studying
neutrino and anti-neutrino interactions with a high intensity, high energy
sign-selected beam.  
Its primary goal is to measure the
electroweak mixing angle $\sin^{2} \theta _{W}$; however, the
high-statistics nature makes it possible to study the rarer processes
described above. At low (zero) hadronic energy the dominant deep-inelastic
scattering (DIS) processes fall off and additional channels such as 
diffractive $D_S^\pm$/$D_S^{*\pm}$ production~\cite{bib:chorus}, 
neutrino trident~\cite{bib:trid1,bib:trid2}, and diffractive 
$J/\psi$~\cite {bib:cdhsjpsi} become more important. This is the first
analysis to study the inclusive production of all processes in the final
state with low hadronic energy and two muons.

The next section will describe the NuTeV experiment while 
Section~\ref{sec:cuts} details the data selection.  Predictions for
the sources of low-$E_{HAD}$ two muons events are given in 
Section~\ref{sec:sources}.  Section~\ref{sec:results} details the
charged- and neutral-current analyses and conclusions are stated
in the final section.

\section{The {NuTeV} Experiment \label{subsec:labe}}

The NuTeV experiment ran at Fermilab using the refurbished Lab E detector~ 
\cite{bib:labe} and $\nu _{\mu }$ and $\bar{\nu}_{\mu }$ beams provided by
the newly installed Sign-Selected Quadrupole Train (SSQT)~\cite{bib:ssqt}.
The SSQT has the capability of selecting either muon neutrino or 
muon anti-neutrino
beams while leaving $\lesssim 2\times 10^{-3}$ of  the anti-selected type.

The Lab E detector, located 1.5 km downstream of the primary target,
consists of a target/calorimeter followed by a toroid spectrometer (Fig.~\ref
{fig:labe}). The calorimeter is composed of 42 segments 
of four steel plates, two liquid scintillator counters (SC) and one drift 
chamber (DC) (Fig.~\ref{fig:calseg}). The steel provides mass 
for the neutrino target,

\noindent
the SC measure the longitudinal vertex and energy deposition ($E$), and 
the DC are used to find the transverse vertex 

and reconstruct downstream tracks.
Because of the high density of the target, only muons deeply penetrate the
calorimeter, and all other particles create a hadronic or electromagnetic
shower near the interaction vertex.

The toroid spectrometer, located immediately downstream of the calorimeter,
focuses muons from the primary charged-current vertex given the type of beam
($\nu_\mu$ \ or $\bar{\nu}_\mu$), measuring both charge and
momentum ($p$) of muons with $p\geq 5$ GeV/c which enter the spectrometer. It is
also possible to measure the momentum of a subset of muons with 
$5\leq p\leq 15$ GeV/c which
range out in the calorimeter, and to place a lower bound on the momentum of
muons which exit the side of the calorimeter.

The NuTeV detector was calibrated with a separate beam of hadrons, muons, or
electrons throughout the running period of the experiment. Hadronic and muon
energy scales for the calorimeter were determined to 0.43$\%$ and $1.0\%$
respectively over the energy range $5-200$ GeV~\cite{bib:nutevnim}. Muon
momentum measurement in the spectrometer is limited by multiple Coulomb
scattering to $\Delta p/p=0.11$, and the sampling-dominated hadronic
resolution was approximately $\Delta E/E=0.86/\sqrt{E}$.

\section{Event Selection \label{sec:cuts}}

\label{sec:cuts}Analyses presented here use the full NuTeV data sample 
from the 1996-1997
fixed-target run corresponding to $1.3\times 10^{18}$  protons on target
(POT) for neutrino running and $1.6\times 10^{18}$ protons on target for
anti-neutrino running.  The size of various event samples from NuTeV
are listed in Table~\ref{tab:nevents}.

Events were selected for the low-$E_{HAD}$ two-muon analysis based on the
following criteria:

\begin{itemize}
\item  The event vertex was required to be at least 25 cm  from any side, 40
cm steel-equivalent from the upstream end and 1.4 m steel-equivalent from
the downstream end of the calorimeter.

\item  At least two muons were required to be found and fitted by the
tracking code. At least one of these had to be reconstructed in the toroid
spectrometer with a momentum greater than 9 GeV/c.
\end{itemize}

\begin{table}[tbp]                                                              
\caption{NuTeV event samples.}                                                  
\label{tab:nevents}                                                             
\begin{tabular}{ccccc}                                                          
& $\#$ of events & $\#$ of events \\                                            
& ($\nu$) & ($\bar{\nu}$) \\                                                    
\hline Single muon (all $E_{HAD}$) & $1.3\times10^6$ & $0.46\times10^6$ \\      
Two muons (all $E_{HAD}$) & 4300 & 1300 \\                                      
Single muon ($E_{HAD} < 3$ GeV) & $0.10\times10^6$ & $0.06\times10^6$ \\        
Two muons ($E_{HAD} < 3$ GeV) & 33 & 21                                         
\end{tabular}                                                                   
\end{table}     

\begin{itemize}
\item  The second muon's momentum could be obtained by either toroid
spectrometer or range information; its momentum was required to be at least
5 GeV/c.

\item  Either the calorimeter drift chamber or scintillator counter signals
had to
be consistent with passage of two muons over five drift chambers or seven 
scintillator counters starting at the event vertex.
\end{itemize}

\section{Low Hadronic Energy, Two Muon Sources \label{sec:sources}}

\label{sec:sources}The analysis strategy consists of comparing data to a
model comprised of all known sources of events with two muons and small
hadronic energy. Kinematic distributions were generated according to
electroweak theory for neutrino trident production, leading order
quark-parton model predictions for DIS feed-down, and vector meson
dominance (VMD) or partially conserved axial current (PCAC) models extended
to four flavors for diffractive charm production. Detector response was
modeled using a GEANT-based Monte Carlo (GEANT 3.21), with simulated 
events processed in
an identical fashion as data. The following sections describe the various
sources and predictions for their rates.

\subsection{DIS Charm Production}

\begin{figure}[tbp]                                                             
 \begin{minipage}{4.2cm}                                                        
  \psfig{figure=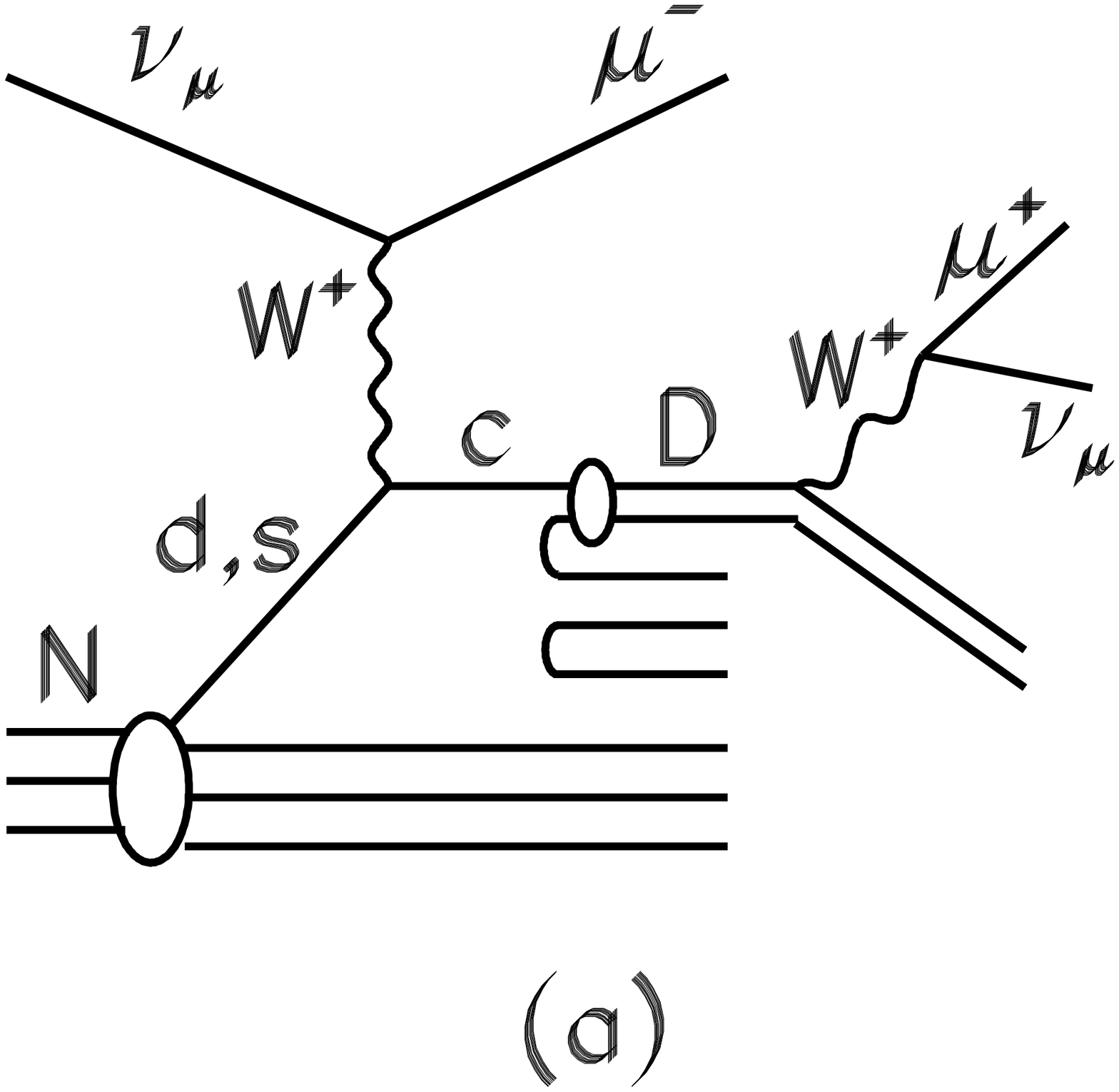,width=4.1cm}                                   
 \end{minipage}                                                                 
 \begin{minipage}{4.2cm}                                                        
  \psfig{figure=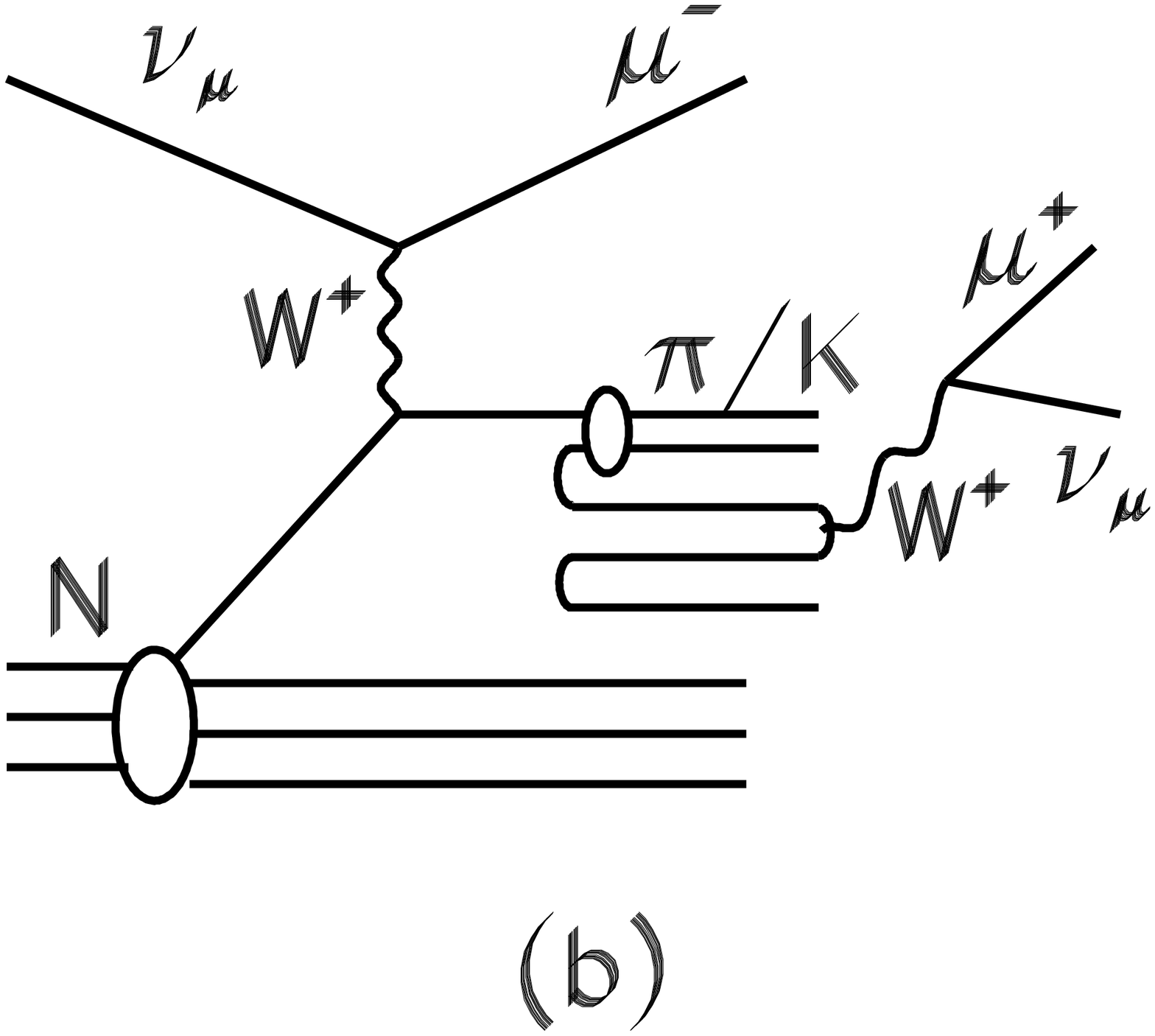,width=4.1cm}                                     
 \end{minipage}                                                                 
\caption{The Feynman diagrams for DIS prod\-uction of two-muon events: (a)      
DIS charged-current production of charm with a semi-muonic decay; (b) DIS       
charged-current production with a $\pi/K$ decay in the hadronic                 
shower.}                                                                        
\label{fig:feyn_dis}                                                            
\end{figure}

Neutrino DIS produces two major sources of two-muon events. Charged-current
charm production can result in a second muon from the decay of the charmed
meson (Fig.~\ref{fig:feyn_dis}(a)). Also, during charged-current
interactions, a $\pi/K$ \ in the hadronic shower can decay to a muon
and a neutrino before interacting (Fig.~\ref{fig:feyn_dis}(b)).

Two-muon events from DIS charm were modeled via leading order predictions
using $d$ and $s$ parton distribution functions measured in this and
previous experiments \cite
{bib:moriond99,bib:charmii,bib:bazarko,bib:rabin,bib:cdhs} assuming an
effective charm quark mass $m_{c}=1.32$ GeV/$c^{2}.$ \ Charm quark
fragmentation was treated using 

\begin{figure}[tbp]                                                             
\centerline{\psfig{figure=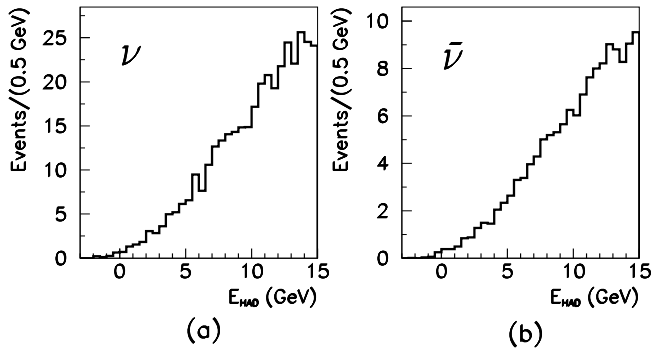,width=\columnwidth}}                  
  \vspace{0.5cm}                                                                
\caption{Distribution of expected hadronic energy (E$_{HAD}$) for DIS           
production of two-muon events (Monte Carlo). This analysis is concerned with    
the contribution near zero.}                                                    
\label{fig:disehad}                                                             
\end{figure}         

\noindent
the Collins-Spiller scheme~\cite
{bib:collins-spiller} with fragmentation parameter $\epsilon =0.93$ and 
$D^{0}:D^{+}:\Lambda _{C}^{+}$ production ratios measured in Fermilab E531~
\cite{bib:e531} and corrected for effects of $D_{S}^{+}$ production~\cite
{bib:bolton-frag}. Charmed hadron decay was modeled using the best
available data as summarized by the Particle Data Group~\cite{bib:pdg}, with
particular attention given to purely leptonic decays.

Two muon events from $\pi /K$ decay in the hadron showers of ordinary
charged-current (CC) events were modeled using measured parameterizations
\cite{bib:sandler} from a previous experiment (FNAL E744/770) which ran at 
similar neutrino energies and used the same neutrino target.

The expected hadronic energy (E$_{HAD}$) spectrum for DIS 
two-muon Monte Carlo events
is shown in Fig.~\ref{fig:disehad}. While these sources generally contain
significant amounts of hadronic energy, some contribution at low 
E$_{HAD}$ \ is seen. Normalizing to the high-E$_{HAD}$ \ 
(E$_{HAD}$ $>$ 5 GeV) two-muon sample we predict 10.4(4.0) events
with E$_{HAD}$ $<$ 3 GeV in $\nu $($\bar{\nu}$) mode due
to the low hadronic energy tail of the DIS process.

\subsection{Neutrino Tridents}

\begin{figure}[tbp]                                                             
 \begin{minipage}{4.2cm}                                                        
  {\psfig{figure=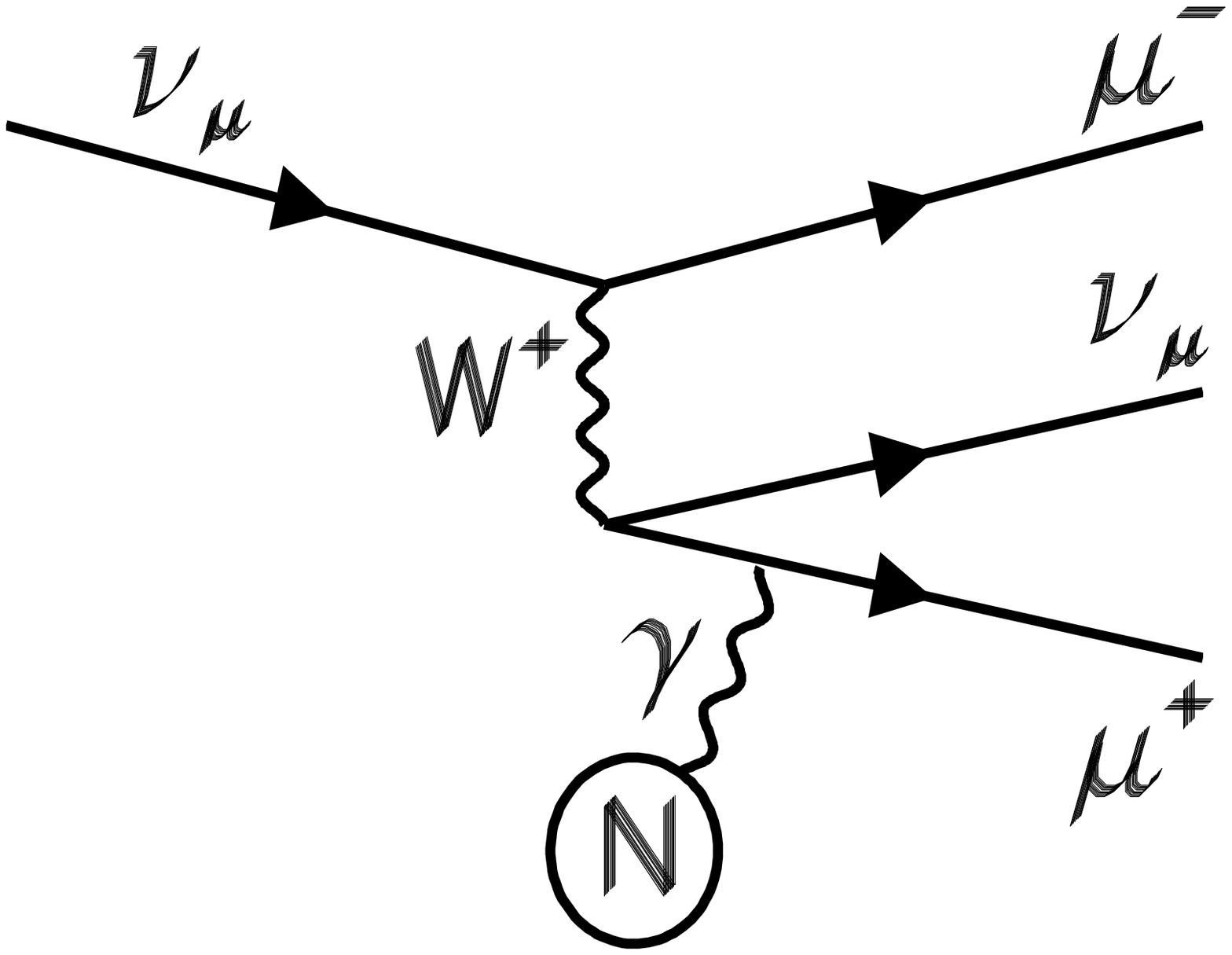,width=4.1cm}}                                    
 \end{minipage}                                                                 
 \begin{minipage}{4.2cm}                                                        
  {\psfig{figure=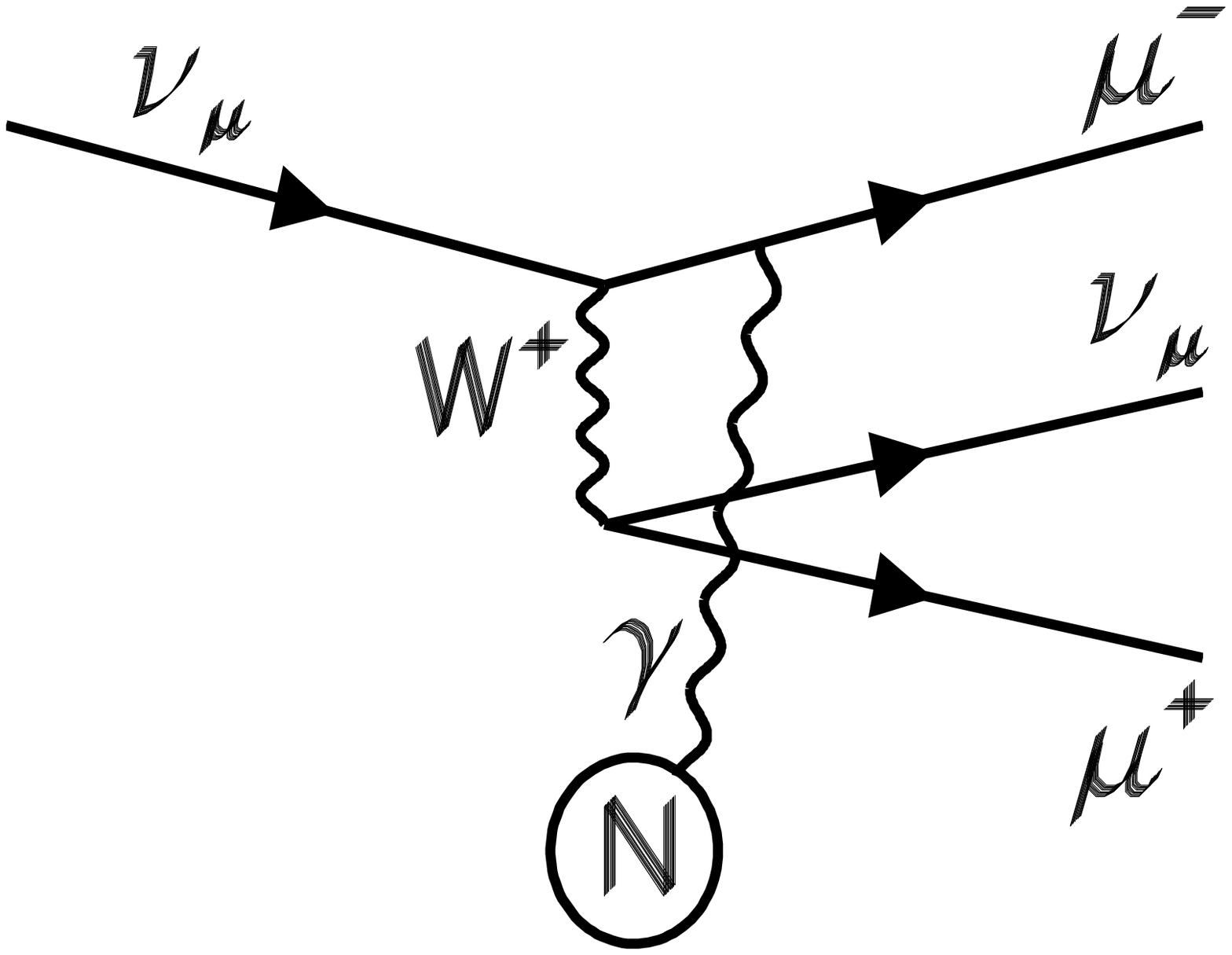,width=4.1cm}}                                    
 \end{minipage}                                                                 
 \vspace{-1cm}                                                                  
                                                                                
 \hspace{4.cm} (a)                                                              
                                                                                
 \hspace{2.cm}                                                                  
 \begin{minipage}{4.2cm}                                                        
  {\psfig{figure=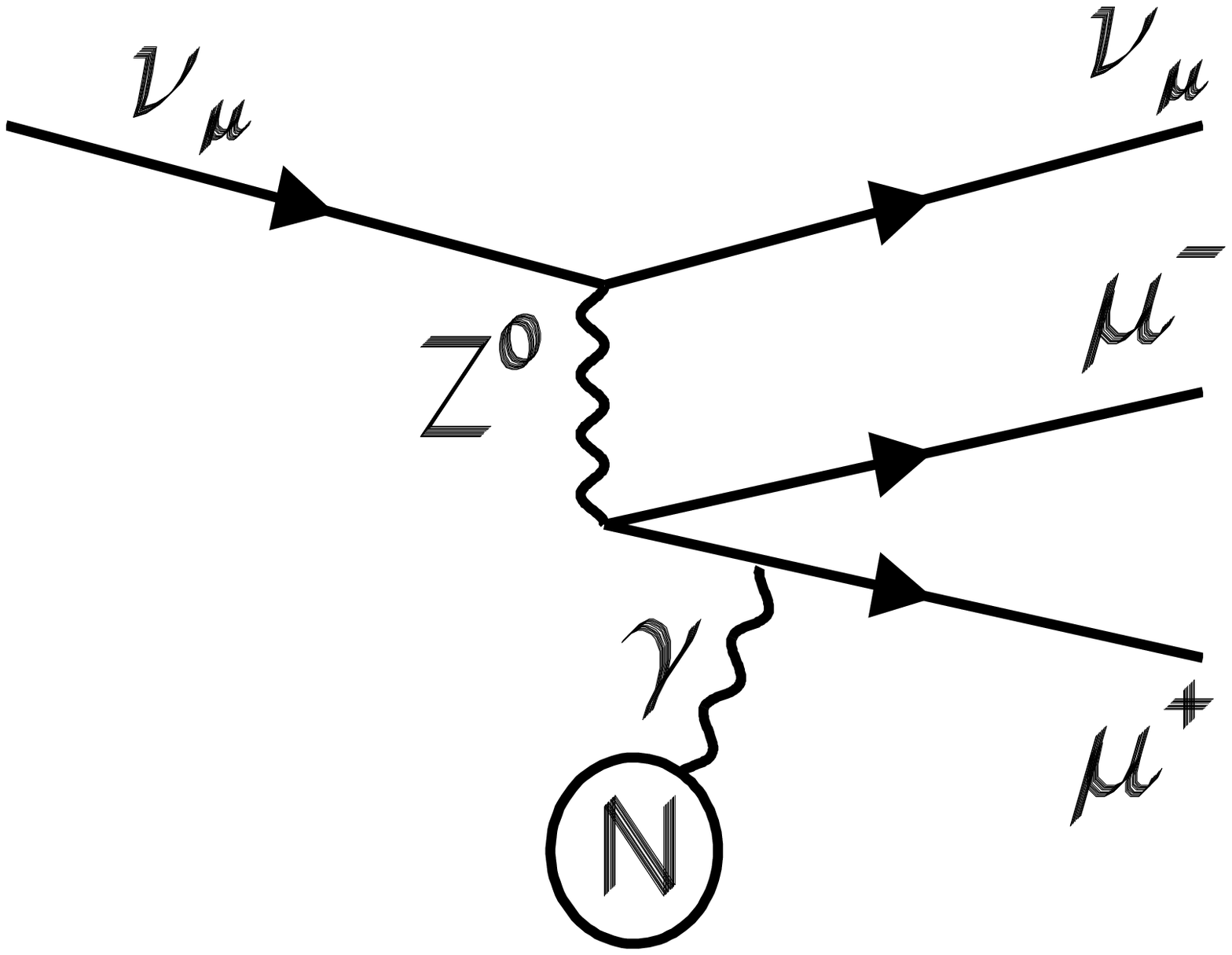,width=4.1cm}}                                    
 \vspace{-0.5cm}                                                                
                                                                                
  \hspace{2.2cm}(b)                                                             
 \end{minipage}                                                                 
  \vspace{0.5cm}                                                                
\caption{The Feynman diagrams for neutrino trident production: (a)              
charged-current production; (b) neutral-current production.}                    
\label{fig:feyn_trid}                                                           
\end{figure}

Neutrino trident production is a purely electroweak process in which
interference between the charged- and neutral-current diagrams causes a 
$40\%$ decrease (from V--A) in the total cross-section~\cite
{bib:trid1,bib:trid2,bib:intbosIII}. Feynman diagrams are shown in Fig.~\ref
{fig:feyn_trid}. The small ($\sim $ $10^{-4}$\ fb/nucleon) cross-section has
limited the observation until recent years~\cite
{bib:ccfrtrid,bib:charmiitrid,bib:charmtrid}.

For this analysis, the Monte Carlo was generated using the full matrix
element with $W$-$Z$ interference~\cite{bib:fuji1,bib:fuji2} including
contributions from both coherent nuclear 
and incoherent nucleon scattering.
This procedure incorporated all possible kinematic correlations between the
two muons and represents an improvement over previous methods\cite
{bib:ccfrtrid,bib:icheptrid}. In particular, there is a strong correlation
between the energies of the two muons which is very important to the
acceptance: when one muon's momentum is high, the other is preferentially
very low.

\begin{figure}[tbp]                                                             
 \begin{minipage}{4.2cm}                                                        
  {\psfig{figure=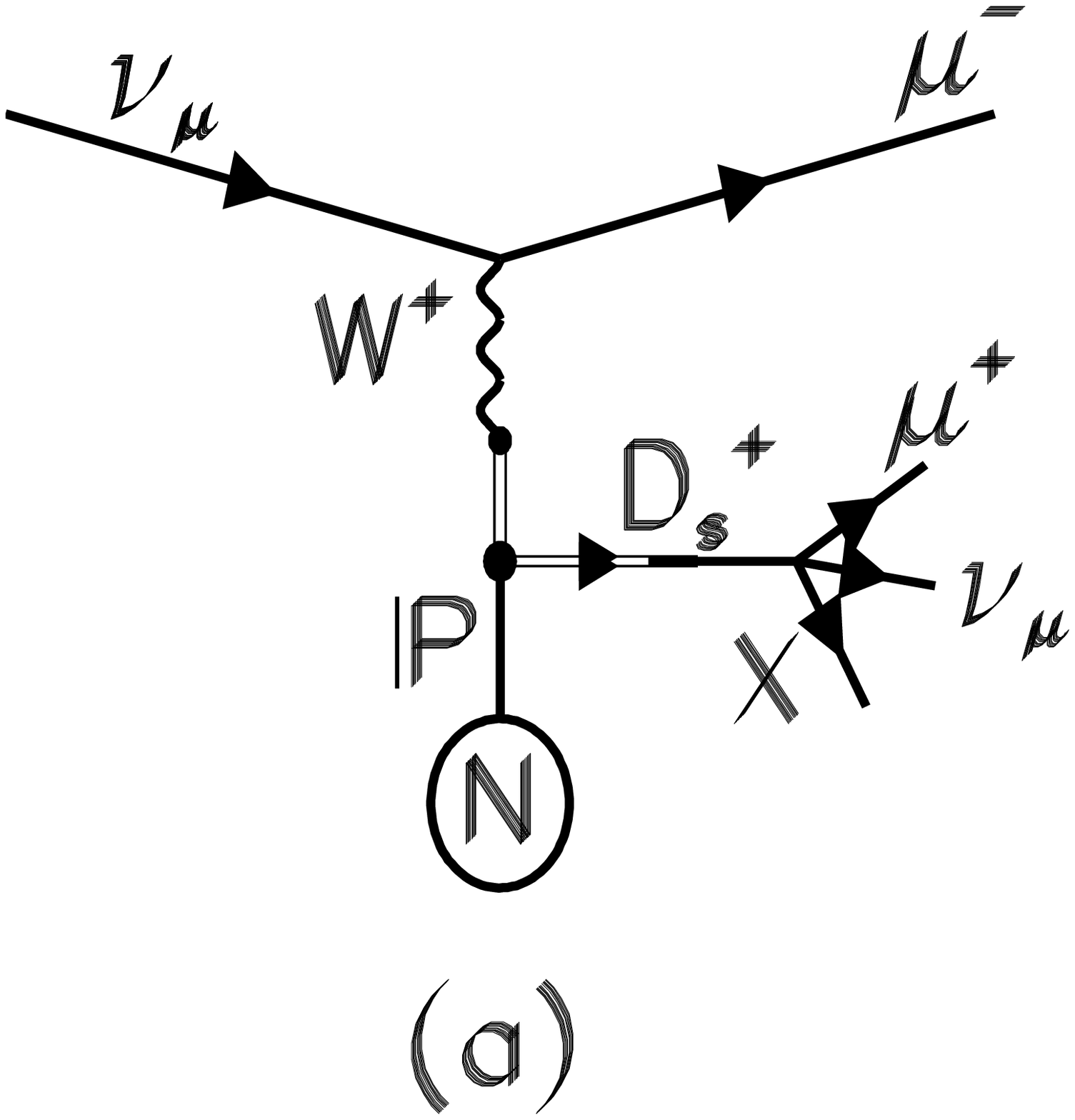,width=4.1cm}}                                       
 \end{minipage}                                                                 
 \begin{minipage}{4.2cm}                                                        
  {\psfig{figure=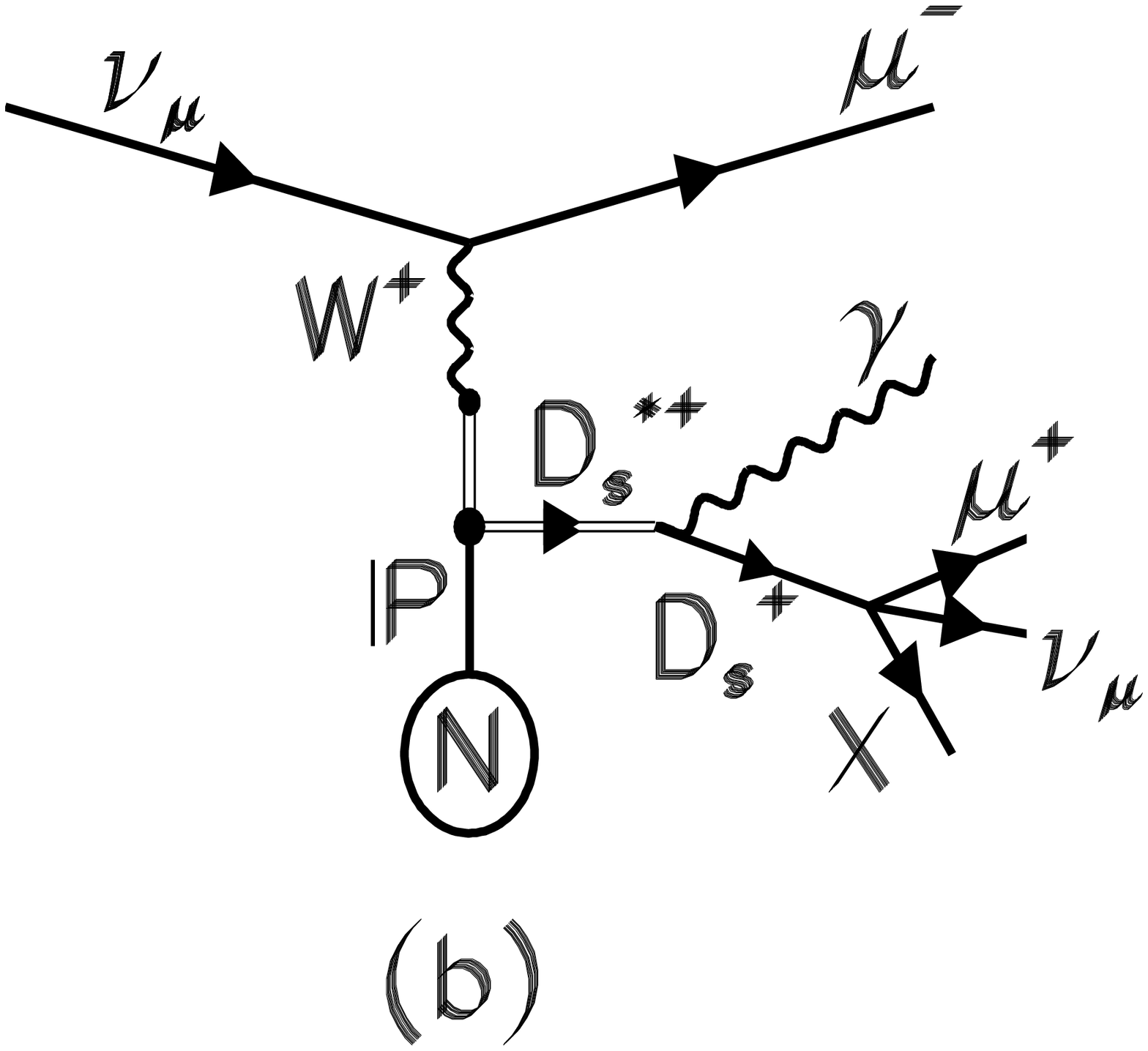,width=4.1cm}}                                   
 \end{minipage}                                                                 
\caption{The Feynman diagrams for (a) $D_S^\pm$ and (b)                         
$D_S^{*\pm}$ production.}                                                       
\label{fig:feyn_ds}                                                             
\end{figure}

\ The lengthy expression for the cross-section can be found in the
references~\cite{bib:trid1,bib:trid2,bib:intbosIII,bib:fuji1,bib:fuji2}. 
The rate depends on the electroweak mixing angle, which we set
to $\sin ^{2}\theta _{W}=0.2222$, and, weakly, on nuclear and nucleon form
factors. Standard dipole parameterizations were used for the latter with a
vector pole mass $m_{V}^{2}=0.71$ GeV$^{2}$/c$^{4}$. The nuclear form factor
for iron was calculated assuming a Fermi charge density function for iron
with nuclear size parameter $c=3.9$ fm and thickness parameter $b=0.55$ fm$.$

NuTeV should observe 4.8(2.2) neutrino trident events in $\nu$($\bar{\nu}$) 
mode from both coherent and incoherent production.

\subsection{$D_S^\pm$/$D_S^{*\pm}$ Production}

Diffractive production of $D_S^\pm$($D_S^{*\pm}$) (Fig.~\ref{fig:feyn_ds}) 
has been observed in previous experiments~\cite{bib:asratyan,bib:chorus} 
and contributes to the low-E$_{HAD}$ \ 
two-muon sample whenever the $D_{S}$ meson decays either to all-lepton final
states $\left( D_{S}^{+}\rightarrow \tau ^{+}\nu _{\tau },\tau
^{+}\rightarrow \mu ^{+}\nu _{\mu }\bar{\nu}_{\tau }\text{ or }
D_{S}^{+}\rightarrow \mu ^{+}\nu _{\mu }\right) $ or to final states with
small hadronic energy $\left( D_{S}^{+}\rightarrow \mu ^{+}\nu _{\mu }X_{
\text{SOFT}}\right)$. 
The latter case was modeled by assuming that the 
$D_{S}^{+}$ semi-muonic decay rate was saturated by the channels 
$D_{S}^{+}\rightarrow \phi \mu ^{+}\nu _{\mu }$, $\eta \mu ^{+}\nu _{\mu }$, 
$\eta ^{\prime }\mu ^{+}\nu _{\mu }$ in the measured proportions summarized
by the Particle Data Group~\cite{bib:pdg}. Only contributions from coherent
diffractive production were considered. Incoherent production of $D_{S}^{+}$(
$D_S^{*\pm}$) from $\nu N$ scattering is already included in the inclusive
DIS rate.

Vector $D_{S}^{\ast }$ production was modeled assuming a VMD type mechanism
with cross-section given by 
\begin{eqnarray}
\frac{d^3\sigma(\nu_\mu N \rightarrow \mu D_{S}^{\ast } N)}{dQ^2 d\nu dt} &
= & \frac{Q^2 \nu}{g_\rho^2 E^2} \ \frac{M_{D_{S}^{\ast }}^2}{(Q^2 +
M_{D_{S}^{\ast }}^2)^2} \nonumber \\
 & & \times \frac{1}{(1 - \epsilon)} \ \frac{e^{-bt}}{b},
\end{eqnarray}
where $Q^2$ and $\nu$ are the momentum and energy transfer to the $D_S^\ast$
, $t$ is the square of the momentum transfer to the nucleus, $g_\rho$ is 
the $\rho$ coupling constant, $M_{D_{S}^{\ast}}$ is the mass of the
$D_{S}^{\ast}$ meson, $E$ is the incoming neutrino energy, $\epsilon$ is
virtual W polarization ($\epsilon = (4E(E - \nu) - Q^2)/(4E(E - \nu) + Q^2 +
2\nu^2)$), and $b$ is the slope of the distribution of momentum transfer 
squared ($t$)
to the nucleus(nucleon) ($b$ = 3 incoherent; $b$ = 145 coherent). 
A check was performed which showed the NuTeV experiment is insensitive to
effects of the virtual-$W$ polarization on the $D_S^{\ast}$ decay.
The overall
normalization was left to be determined by the data. To set the scale, the
normalization was also obtained in the $SU(4)$-flavor limit by comparing to
the measured cross-section for diffractive $\rho ^{+}$ production in
neutrino scattering. 

The number of events expected by NuTeV is normalized to the observed
inclusive two-muon data with E$_{HAD}$ $>$ 5 GeV 
\begin{equation}
N_{M}=\frac{\int \Phi \times \sigma _{M}(\mbox{$E_\nu$})\times \epsilon (
\mbox{$E_\nu$})\times BF_{M} \ d\Phi}{\int \Phi \times \sigma _{\mbox{$\mu\mu$}
}(\mbox{$E_\nu$})\times \epsilon (\mbox{$E_\nu$})\times BF_{\mbox{$\mu\mu$}
} \ d\Phi }\times N_{\mbox{$\mu\mu$}}  \label{eq:lvmnorm},
\end{equation}
where $M$ refers to the meson being produced, $\mu\mu$ \
refers to DIS two-muon data, $\Phi$ is the NuTeV flux, $\sigma$ is the
cross-section, $E_\nu$ is the incident neutrino energy, $\epsilon$ is the 
detection efficiency, BF is the (semi-)muonic branching
fraction, and $N_{\mu\mu}$ is the number of DIS two-muon events
observed in the NuTeV data. 
This procedure predicts 33.0(13.6) observed $D_S^{\ast}$ events for 
$\nu$($\bar{\nu}$) mode in NuTeV.

The contribution of pseudoscalar $D_{S}^{+}$ decay was estimated using PCAC
formulas~\cite{bib:e632,bib:kopel} adapted for charm: 
\begin{eqnarray}
\frac{d^3\sigma(\nu_\mu N \rightarrow \mu D_{S} N)}{dQ^2 d\nu dt} & = & 
\frac{\nu}{E^2} \frac{\epsilon}{(1 - \epsilon)} \ \frac{M_{D_{S1}}^4}{(Q^2 +
M_{D_{S1}}^2)^2} \nonumber \\
& & \times f_{D_S}^2 \ \frac{e^{-b t}}{b},
\end{eqnarray}
\begin{figure}[tbp]                                                             
\centerline{\psfig{figure=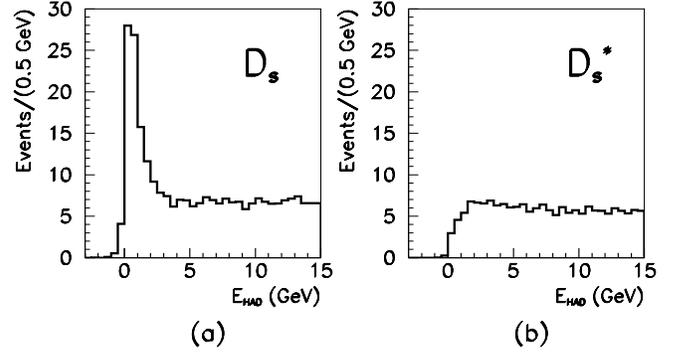,width=\columnwidth}}                   
\caption{The E$_{HAD}$ distributions for (a) $D_S^\pm$ and                      
(b) $D_S^{*\pm}$ Monte Carlo.}                                                  
\label{fig:ehad_ds}                                                             
\end{figure}  

\noindent
where $M_{D_{S1}}$ is the mass of the $D_{S1}$ meson and $f_{D_S}$ is the 
$D_S$ decay constant ($f_{D_S}$ = 0.31 GeV). 102.9(48.2) events are predicted
for $\nu$($\bar{\nu}$) mode in the SU(4) flavor limit.

Figure~\ref{fig:ehad_ds} shows the E$_{HAD}$ distributions
predicted by the Monte Carlo. The $\tau $ decay mode results in the peak at 
E$_{HAD}$ = 0 for $D_S^\pm$ production while the extra
photon from the $D_S^{*\pm}$ decay washes the peak out.

\subsection{Other Sources}

Several other contributions to low-$E_{HAD}$ two-muon states contribute at
the sub-event level. These include $\mu ^{+}\mu ^{-}$ decays from
neutral-current (NC) diffractively produced $J/\psi $, $\mu ^{+}\mu ^{-}$
decays from NC diffractively produced vector mesons, CC diffractively
produced $\pi ^{+}$ in which the pion decays, $\pi ^{+}$ from CC baryon
resonance production followed by pion decay, and quasi-elastic CC scattering
where pattern recognition errors split the outgoing muon track into two.

\subsubsection{NC Diffractive Vector Meson Sources}

\begin{figure}[tbp]
\centerline{\psfig{figure=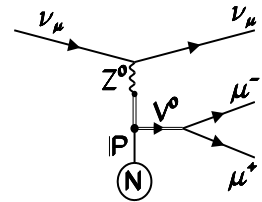,width=4cm}}                               
\caption{The Feynman diagram for neutral-current diffractive production of      
light vector mesons (V$^0$). Here V$^0$ can be a $\rho^0$, $\omega$, $\phi$     
or $J/\psi$ meson and N represents either a nucleus (coherent) or               
nucleon (incoherent).}                                                          
\label{fig:feyn_lvm}                                                            
\end{figure}

Diffractive production of vector mesons is an important process in
photoproduction experiments. A similar process is available in the weak
sector with the substitution of a $Z^{0}$ for the photon (Fig.~\ref
{fig:feyn_lvm}). Vector mesons
which can decay to two muons include $\rho
^{0}$, $\omega $, $\phi $ and $J/\psi$.

The VMD model is used to predict the expected number of vector mesons.
Cross-sections are normalized to the Fermilab E632 measurement of neutrino
production of single $\rho ^{\pm }$~\cite{bib:e632} with the neutral-current
(NC) and charged-current (CC) cross-sections related by Pumplin's procedure~
\cite{bib:pumplin}. The $J/\psi $ estimate is given in the $SU\left(
4\right) $ flavor limit for comparison purposes.

Table~\ref{tab:lvm} lists the number of vector meson events predicted to be
observed in the NuTeV data. All sources except $J/\psi$ production
are expected to be small, with the suppression largely attributable to the very
small $\mu ^{+}\mu ^{-}$ branching fractions for vector mesons. Section~\ref
{sec:jpsimeas} describes a measurement of the $J/\psi$ signal from
diffractive production.

\subsubsection{Sources from Pion Decay or Mis-reconstruction}

\begin{figure}[tbp]                                                             
 \begin{minipage}{4.2cm}                                                        
  {\psfig{figure=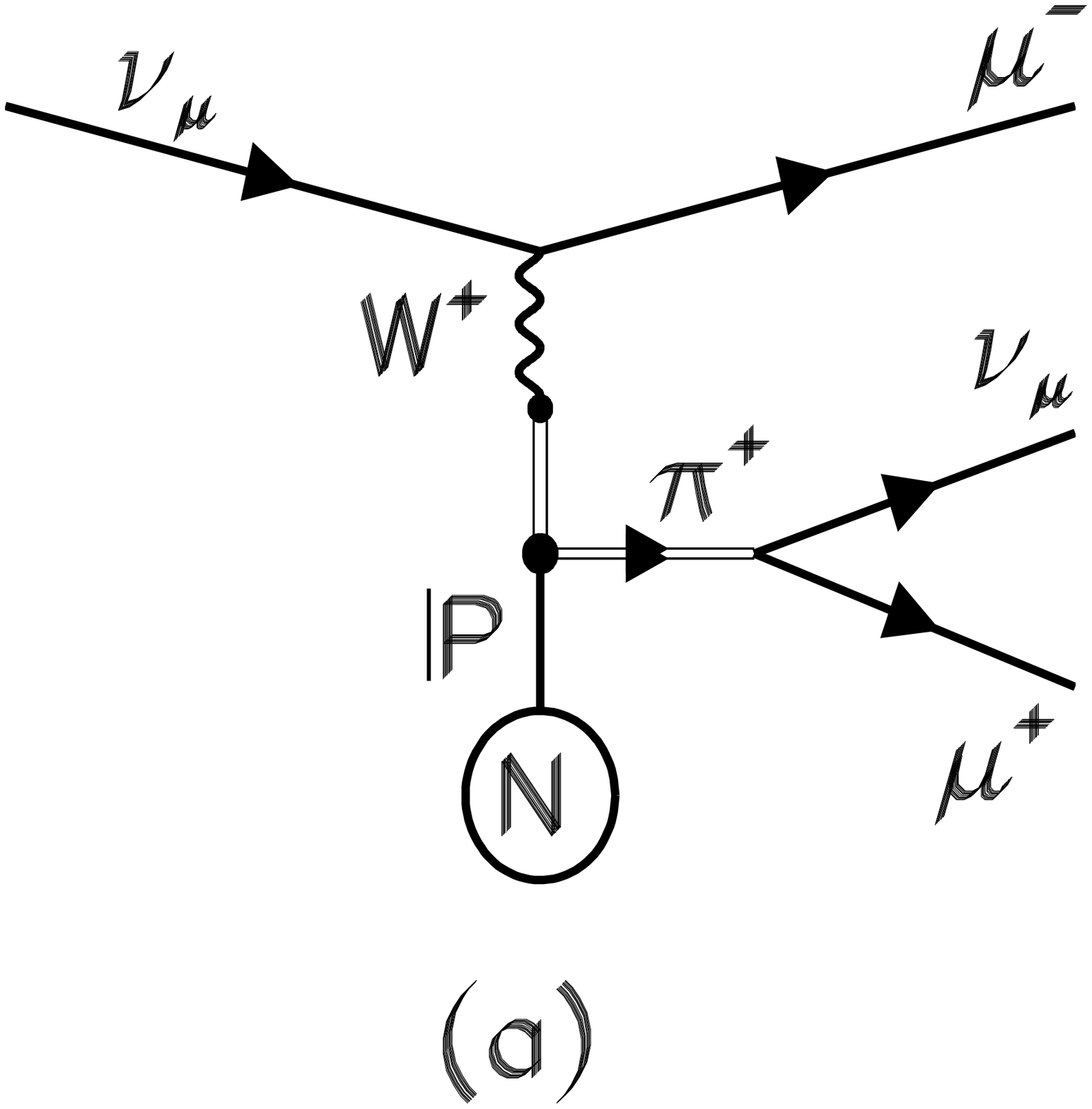,width=4.1cm}}                                     
 \end{minipage}                                                                 
 \begin{minipage}{4.0cm}                                                        
  \vspace{-0.5cm}                                                               
                                                                                
  \hfill                                                                        
  {\psfig{figure=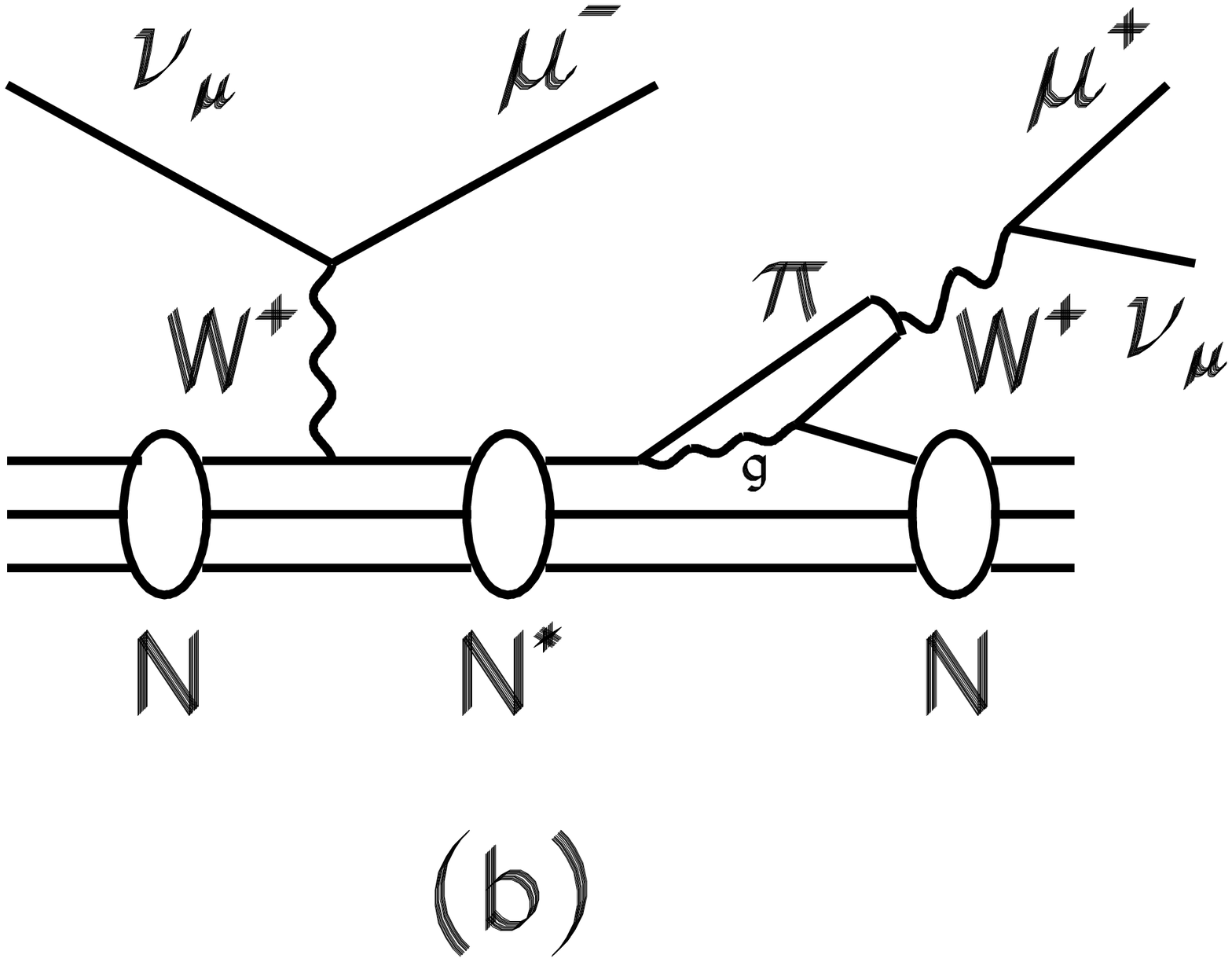,width=3.9cm}}                                    
 \end{minipage}                                                                 
  \vspace{0.5cm}                                                                
\caption{The Feynman diagrams for single $\protect\pi^\pm$ production: (a)      
diffractive and (b) resonance. }                                                
\label{fig:feyn_pion}                                                           
\end{figure}

Single pions produced in NuTeV have two primary sources: diffractive and
resonance production (Fig.~\ref{fig:feyn_pion}). The density of the NuTeV
detector limits the contribution of both sources to the two-muon sample
since the pion must decay prior to interacting. Diffractive production is
predicted using the PCAC model normalized to the measured cross-section~\cite
{bib:e632} resulting in 0.08(0.02) estimated events in $\nu$($\bar{\nu}$) 
mode. Resonance production can result in a decay of a
final state pion with little visible hadronic energy. This is calculated
with a model from Rein and Sehgal~\cite{bib:rein} and predicts $<$
0.1 events in either mode.

\begin{figure}[tbp]                                                             
  \centerline{\psfig{figure=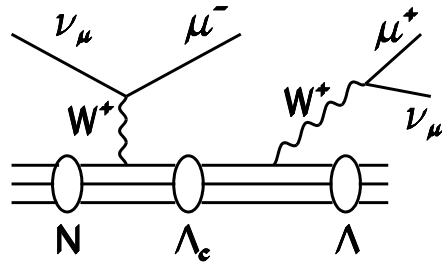,width=6cm}}                         
  \vspace{0.5cm}                                                                
\caption{The Feynman diagram for diffractive $\Lambda_C^\pm$ production.}       
\label{fig:feyn_lambdac}                                                        
\end{figure}

Quasi-elastic production of $\Lambda_C^\pm$ events 
(Fig.~\ref{fig:feyn_lambdac}) has also been considered~\cite{bib:shrock}. 
The muon from the $\Lambda_C^\pm$ decay is of very low momentum resulting in
extremely low acceptance. We predict $<$0.1 events in either mode.

The final source considered was leakage from low-E$_{HAD}$ single
muon events due to mis-reconstruction. Contributions from this class of
events was greatly reduced by requiring that the scintillator counters or
drift chambers were consistent with
two muons over a minimum length. The primary remaining source of these
events was quasi-elastic scattering where multiple coulomb scattering and 
drift chamber inefficiencies 
caused the pattern recognition software to find a spurious
second track in the event. This contribution was estimated using
straight-through muons which result from 

\begin{table}[tbp]                                                              
\caption{VMD predictions for the expected number of diffractively produced      
light vector mesons which decay to two muons.}                                  
\label{tab:lvm}                                                                 
\begin{tabular}{ccccc}                                                          
& $\#$ of events & $\#$ of events & $\#$ of events & $\#$ of events \\          
& (coherent) & (incoherent) & (coherent) & (incoherent) \\                      
& ($\nu$) & ($\nu$) & ($\bar{\nu}$) & ($\bar{\nu}$) \\                          
\hline $\rho^0$ & 0.0074 & 0.0039 & 0.0037 & 0.0019 \\                          
$\omega$ & 0.023 & 0.016 & 0.011 & 0.008 \\                                     
$\phi$ & 0.036 & 0.022 & 0.018 & 0.011 \\                                       
$J/\psi$ & 4.88 & 17.1 & 1.86 & 7.76                                            
\end{tabular}                                                                   
\end{table}    

\noindent
charged-current interactions
upstream of the detector that enter the calorimeter. A random longitudinal
event vertex was chosen, and all event information upstream of this point
was discarded, leaving a topology identical to a quasi-elastic event. These
straight-through muons were analyzed in the same way as the neutrino data sample
and the results normalized to the observed number of single muon
quasi-elastic events.
The resulting predictions are 0.75(0.25) events in $\nu$($\bar{\nu}$) mode.

A summary of the predicted number of events from each source is shown in
Table~\ref{tab:source_sum}. The remainder of this paper describes the
measurement of the largest sources: DIS, neutrino tridents, $D_S^\pm$/
$D_S^{*\pm}$, and $J/\psi$.

\section{Results} \label{sec:results}

\subsection{Charged-Current Analysis}

\label{sec:measure}
\begin{figure}[tbp]                                                             
  \centerline{\psfig{figure=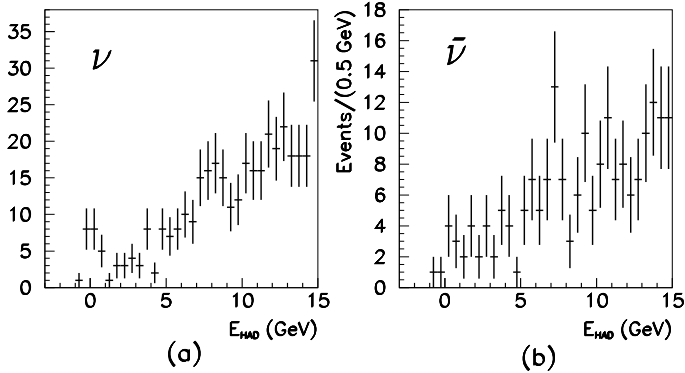,width=\columnwidth}}               
\caption{The hadronic energy distribution for the two-muon data.                
The first twelve bins contribute to                                             
figures~\ref{fig:mmumu} and~\ref{fig:deltaphi}. }                               
\label{fig:ehcc3data}                                                           
\end{figure}

\begin{figure}[tbp]                                                             
  \centerline{\psfig{figure=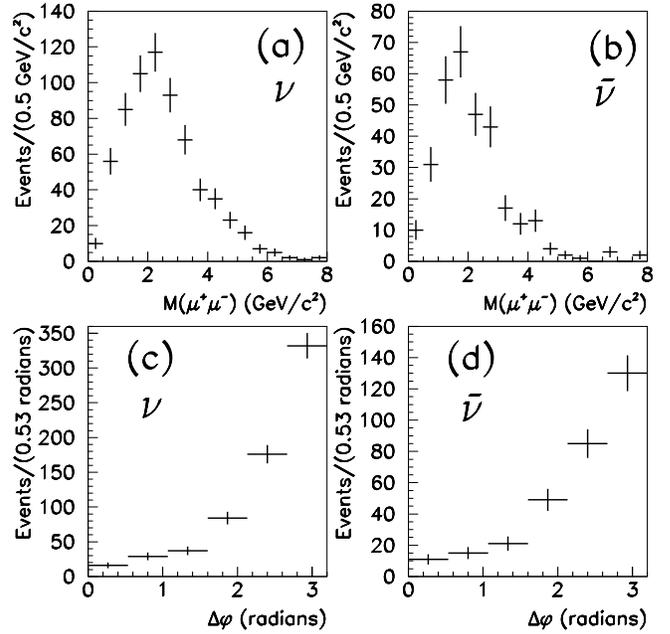,width=\columnwidth}}                
\caption{Distributions for the two-muon NuTeV events with E$_{HAD} \ <$ 15 GeV. 
The plots are the two-muon invariant mass ($M_{\mu\mu}$) for                    
$\nu$ (a) and $\bar{\nu}$ (b) modes and $\Delta \phi$ in $\nu$ (c) and          
$\bar{\nu}$ (d) modes. }                                                        
\label{fig:ehadlt15}                                                            
\end{figure}

\begin{figure}[tbp]
 \vspace{1.5cm}

  \centerline{\psfig{figure=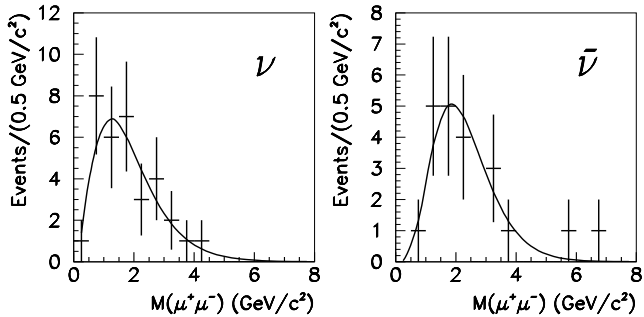,width=\columnwidth}}                   
\caption{The two-muon invariant mass ($M_{\mu\mu}$) for the low-                
E$_{HAD}$ (E$_{HAD}$ $<$ 3 GeV) two-muon data.                                  
The curve shows a fit to the                                                    
distribution $x^\protect\protect\alpha e^{(\protect\beta + \protect\gamma       
x)} $. }                                                                        
\label{fig:mmumu}                                                               
\end{figure}

\begin{figure}[tbp]                                                             
 \vspace{1.5cm}

  \centerline{\psfig{figure=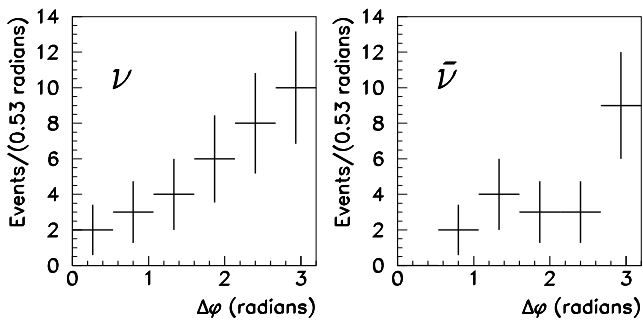,width=\columnwidth}}                
\caption{The $\Delta \protect\phi$ distributions for the low-E$_{HAD}$          
(E$_{HAD}$ $<$ 3 GeV) two-muon data.                                            
$\Delta \protect\phi$ is defined in the text. }                                 
\label{fig:deltaphi}                                                            
\end{figure}

Our charged-current analysis examines three kinematic
variables: the hadronic energy ($E_{HAD}$) up to 15 GeV, 
the two-muon
invariant mass ($M_{\mu \mu }$), and the absolute value of the smallest
difference between the muon azimuthal angles ($\Delta \phi $). Distributions
for these quantities 
as measured in NuTeV data are shown in Figs.~\ref
{fig:ehcc3data}, \ref{fig:ehadlt15}, \ref{fig:mmumu}, 
and~\ref{fig:deltaphi}. \
Since the analysis is primarily concerned 

\begin{minipage}{17.8cm}
\begin{center}
\begin{minipage}{12cm}
\begin{table*}[tbp]                                                             
\caption{Predictions for the number of events expected to be                    
observed by NuTeV in the low-E$_{HAD}$ (E$_{HAD}$ $<$ 3                         
GeV) two-muon sample. }                                                         
\label{tab:source_sum}                                                          
\begin{tabular}{cccc}                                                           
& $\#$ of events & $\#$ of events &  \\                                         
& ($\nu$) & ($\nu$) &  \\                                                       
\hline                                                                          
DIS & 10.4 & 4.0 & Ref.~\cite{bib:moriond99}  \\                                
Neutrino Tridents & 4.8 & 2.2 & Ref.~\cite{bib:trid1,bib:trid2}  \\             
$D_S^\pm$ & 102.9 & 48.2 & Ref.~\cite{bib:asratyan,bib:chorus}  \\              
$D_S^{*\pm}$ & 33.0 & 13.6 & Ref.~\cite{bib:asratyan,bib:chorus}  \\            
Light vector mesons ($\rho^0$, $\omega$, $\phi$) & $<$ 0.12 & $<$ 0.08 & Ref.\cite{bib:e632,bib:pumplin}  \\                                                    
$J/\psi$ (coherent) & 4.88 & 1.86 & Ref.\cite{bib:e632,bib:pumplin} \\          
$J/\psi$ (incoherent) & 17.1 & 7.76 & Ref.\cite{bib:e632,bib:pumplin} \\        
Single $\pi^\pm$ & $<$0.1 & $<$0.1 & Ref.\cite{bib:e632,bib:rein}  \\           
$\Lambda_C^\pm$ & $<$0.1 & $<$0.1 &  Ref.\cite{bib:shrock} \\                   
Mis-identified single muon events & 0.75 & 0.25 &   \\                          
\hline Sum & 174.4 & 78.2 &                                                     
\end{tabular}
\end{table*}                                                                    
\end{minipage}
\end{center}
\end{minipage} 
\vspace{2cm}

\vspace{2cm}

\vspace*{8cm} 

\vspace{1cm}

\noindent
with events that peak at low $
E_{HAD}$, the $M_{\mu \mu }$ (Fig.~\ref{fig:mmumu}) and 
$\Delta \phi$ (Fig.~\ref{fig:deltaphi}) distributions are plotted
with the additional restriction $E_{HAD}\leq 3$ GeV to emphasize the
diffractive region.  However, all events with $E_{HAD}\leq 15$ GeV\ are used
in the subsequent fitting. Thirty-three neutrino and 21 anti-neutrino
two-muon
events were observed with E$_{HAD}$ $<$ 3 GeV. These numbers are
significantly below the sum of the predicted number of events in Table~\ref
{tab:source_sum}, indicating the naive $SU(4)$ flavor predictions for heavy
quark final states are too large. On the other hand, there is a large excess
over the predictions from DIS feed-down and neutrino trident production.

A fit was performed allowing up to four sources of low-E$_{HAD}$ 
two-muon events: DIS, neutrino tridents, diffractive $D_S^\pm$, and
diffractive $D_S^{*\pm}$.  These sources were simulated via Monte
Carlo and normalized to the 
NuTeV inclusive two-muon samples.  Figure~\ref
{fig:mcdists} shows distributions of $E_{HAD}$, $M_{\mu \mu }$, and $\Delta
\phi $ for the four major sources. Deep-inelastic scattering contributions
peak at high $M_{\mu \mu }$, high $E_{HAD}$, and at $\Delta \phi =0$ and $\pi 
$, the latter feature reflecting the essentially two-body final state of
DIS charm production. The neutrino trident signal peaks at low $E_{HAD}$,
low $M_{\mu \mu}$ and is more uniform in $\Delta \phi $. The $D_{S}$ and 
$D_{S}^{\ast}$ distributions are intermediate between DIS and neutrino
tridents, with $D_{S}$ distinguishable from $D_{S}^{\ast }$ in the $E_{HAD}$
distribution due to the decay photon contribution to $E_{HAD}$ for 
$D_{S}^{\ast }$. Diffractive $D_{S}$ and neutrino trident distributions are
very similar, hence their measurements are highly correlated. 

Data and Monte Carlo were binned three-dimensionally in 
$\left( E_{HAD},M_{\mu \mu },\Delta \phi
\right) $ space: 18 bins in $E_{HAD}$ \ (-3 - 15 GeV), 6 in 
$M_{\mu\mu}$ (0 - 6 GeV); \ and 6 in $| \Delta \phi |$ (0
- $\pi $ radians). The Monte Carlo is fit to the data using a
maximum-likelihood technique. The Monte Carlo sets are summed together using
a single normalization factor for each source which was defined to be 1.0
for the level presented in the preceding section. Neutrino and anti-neutrino
modes were fit simultaneously.

\begin{figure}[tbp]                                                             
  \centerline{\psfig{figure=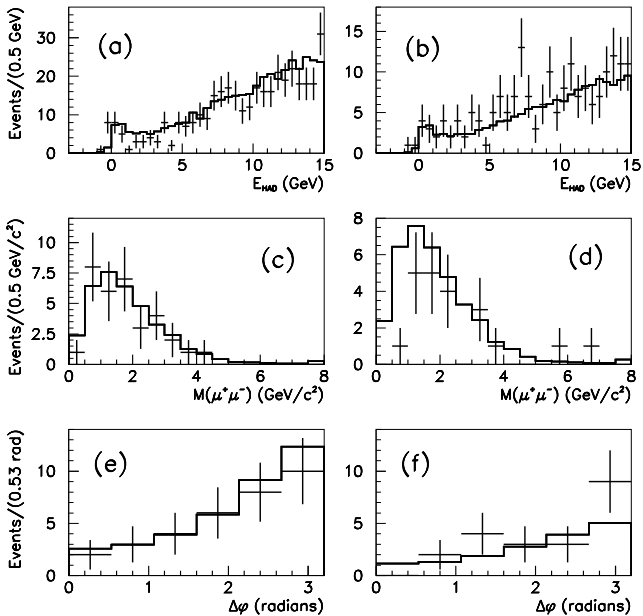,width=\columnwidth}}               
\caption{Comparison of the final result (MC) to the low-E$_{HAD}$               
two-muon data for (a,b) E$_{HAD}$, (c,d) $M_{\mu^+\mu^-}$,                      
(e,f) $\Delta \protect\phi$.  The left side is $\nu$ mode; the right side       
is $\bar{\nu}$ mode.  The $M_{\mu^+\mu^-}$ and $\Delta \protect\phi$            
distributions are for $E_{HAD}$ $<$ 3 GeV. The points represent the data        
while the histogram shows the Monte Carlo.}                                     
\label{fig:lehfitres}                                                           
\end{figure}

The results of the fit are shown in Table~\ref{tab:lehfitresults}. 
Figure~\ref{fig:lehfitres} compares these results to the data and shows 
we are able
to describe the data with the four largest sources. The DIS contribution is
consistent with that expected from the higher E$_{HAD}$ two-muon
analysis~\cite{bib:moriond99}. The neutrino trident contribution is
consistent with the Standard Model prediction, but can not distinguish
between V--A and the Standard Model. This is in contrast to previous analyses
which ruled out V--A but did not consider diffractive sources.
\vspace{2cm}

\begin{table}[tbp]                                                              
\caption{Parameters from the three parameter fit to the low-E$_{HAD}$           
\ two-muon data. }                                                              
\label{tab:lehfitresults}                                                       
\begin{tabular}{lc}                                                             
Parameter & Result \\                                                           
\hline DIS & 0.90                                                               
\raisebox{+0.7ex}{$\begin{array}{c} +0.09 \\ -0.08                              
\end{array}$} \\                                                                
Neutrino Tridents & 0.72                                                        
\raisebox{+0.7ex}{$\begin{array}{c} +1.73 \\ -0.72                              
\end{array}$} \\                                                                
Diffractive Charm ($D_S^\pm$ \ + $D_S^{*\pm}$) & 0.18 %
\raisebox{+0.7ex}{$\begin{array}{c} +0.06 \\ -0.06 \end{array}$}                
\end{tabular}                                                                   
\end{table}    

\newpage
\begin{minipage}{17.8cm}
\begin{center}
\begin{figure*}[tbp]                                                            
 \begin{minipage}{16cm}                                                         
  \centerline{\psfig{figure=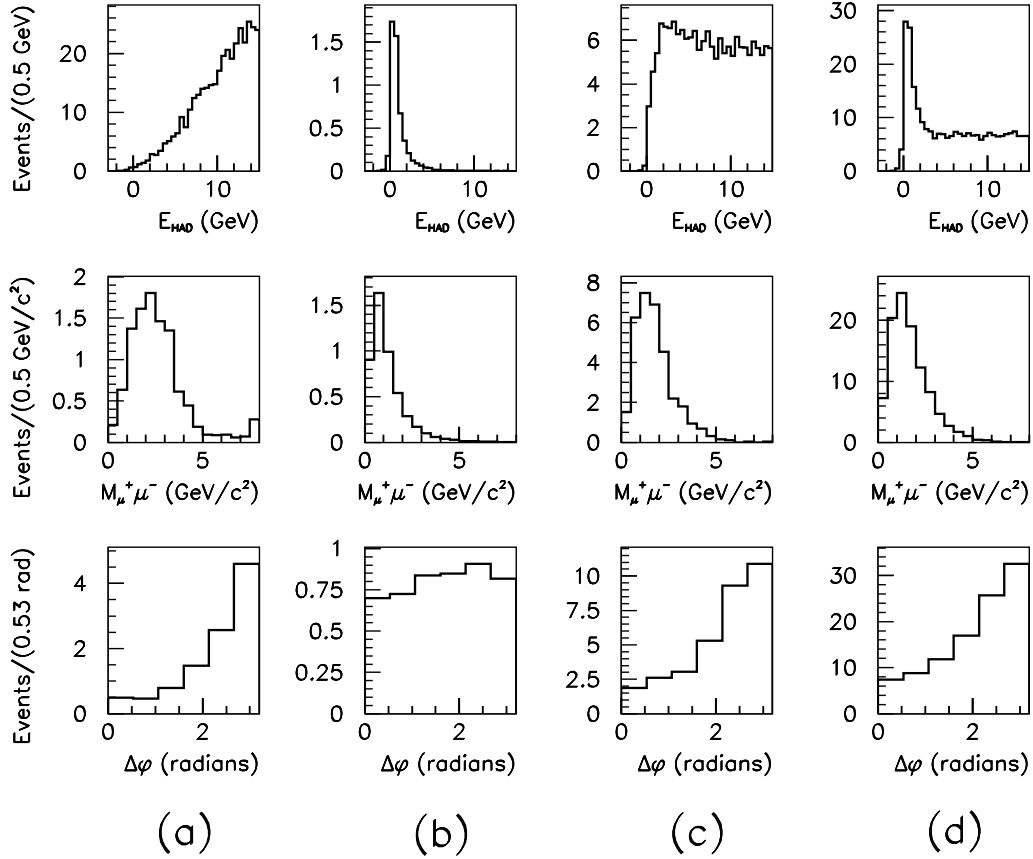,width=14cm}}                         
\caption{Monte Carlo distributions of visible hadronic energy, two muon         
invariant mass and $\Delta \protect\phi$ for the four the four largest          
sources of low-E$_{HAD}$ two-muon events: (a) DIS charm; (b)                    
neutrino tridents; (c) diffractive $D_S^\pm$; (d) diffractive                   
$D_S^{*\pm}$. The invariant mass and $\Delta \protect\phi$                      
distributions are for E$_{HAD}$ $<$ 3 GeV. }                                    
\label{fig:mcdists}                                                             
 \end{minipage}
\end{figure*}
\end{center}
\end{minipage}
\vspace{1cm}

\begin{minipage}{17.6cm}
\begin{center}
 \begin{minipage}{12cm}
\begin{table}[tbp]
\caption{Measured event contributions to the low-E$_{HAD}$ two-muon             
event sample.}                                                                  
\label{tab:leheventresults}                                                     
\begin{tabular}{ccccc}                                                          
& $\#$ of events & $\#$ of events &  &  \\                                      
& ($\nu$) & ($\nu$) &  &  \\                                                    
\hline \\ \vspace{0.2cm}                                                        
$J/\psi$ (90$\%$ CL) & $<$7.5  & $<$5.0 &  &  \\ \vspace{0.2cm}                 
DIS                                                                             
 & 9.4 \raisebox{+0.7ex}{$\begin{array}{c} +0.9 \\ -0.9 \end{array}$}           
 & 3.6 \raisebox{+0.7ex}{$\begin{array}{c} +0.4 \\ -0.4 \end{array}$} &  & \\ \v
space{0.2cm}                                                                    
Neutrino Tridents                                                               
 & 3.5 \raisebox{+0.7ex}{$\begin{array}{c} +8.3 \\ -3.5 \end{array}$}           
 & 1.6 \raisebox{+0.7ex}{$\begin{array}{c} +3.8 \\ -1.6 \end{array}$} &  & \\ \v
space{0.2cm}                                                                    
($D_S^\pm$ + $D_S^{*\pm}$)                                                      
 & 24.5 \raisebox{+0.7ex}{$\begin{array}{c} +8.1 \\ -8.2 \end{array}$}          
 & 11.1 \raisebox{+0.7ex}{$\begin{array}{c} +3.7 \\ -3.7 \end{array}$} &  & \\ \
vspace{0.2cm}                                                                   
Mis-identified single muon events & 0.75 $\pm$ 0.43 & 0.25 $\pm$ 0.25 &  &  \\ \
vspace{0.2cm}         
All other sources (estimate) & $<$0.1 & $<$0.1 &  & \\                          
\end{tabular}                                                                   
\end{table}   
 \end{minipage}
\end{center}
\end{minipage}

\newpage
\vspace*{22cm}
\newpage

The consideration of all sources of low-E$_{HAD}$ two-muon events
allows us to measure diffractive charm production. The $D_S^\pm$
and $D_S^{*\pm}$ sources have been combined in proportion to the
theoretical predictions and a single fit parameter used. This yields
cross-sections of 
\begin{eqnarray*}
\sigma \left( \nu _{\mu }Fe\rightarrow \mu ^{-}(D_{S} \ + \ 
D_{S}^{\ast })Fe\right)
&=&(3.3 \pm 1.1)\text{ fb}/\text{nucleon,}
\end{eqnarray*}
evaluated at $E_\nu$ = 130 GeV using the modified VMD\ and PCAC
predictions to extrapolate in energy under the assumptions $\sigma \left(
\nu _{\mu }Fe\rightarrow \mu ^{-}D_{S}^{\ast +}Fe\right) =\sigma \left( \bar{
\nu}_{\mu }Fe\rightarrow \mu ^{+}D_{S}^{\ast -}Fe\right) $ and $\sigma
\left( \nu _{\mu }Fe\rightarrow \mu ^{-}D_{S}^{+}Fe\right) =\sigma \left( 
\bar{\nu}_{\mu }Fe\rightarrow \mu ^{+}D_{S}^{-}Fe\right) $. A second fit
performed with the neutrino trident parameter fixed to the Standard Model
prediction yielded the consistent results 
$\sigma \left( \nu _{\mu }Fe\rightarrow \mu ^{-}(D_{S} \ + \ D_{S}^{\ast })Fe
\right) =(3.0\pm 0.8)$ fb$/$nucleon at $E_\nu$ =
130 GeV. The quoted errors are completely dominated by statistics.  
This result assumes an isotropic $D_S^{\ast}$ decay.  Studies showed
effects of a possible $D_S^{\ast}$ polarization to be small.  The
largest change, corresponding to nearly complete longitudinal
polarization, lowered $\sigma(D_S \ + \ D_S^{\ast}$) by 
0.4 fb/nucleon.

Previously, the Big Bubble Chamber Neutrino Collaboration combined various
data samples to measure the diffractive rate of charmed strange mesons (
$D_S^\pm$ + $D_S^{*\pm}$) per charged-current $\nu I$ ($I$
is an isoscalar target) interaction~\cite{bib:asratyan}. They measured a
rate of (2.8 $\pm $ 1.1) $\times $ 10$^{-3}$. The observation of 
$D_S^{*\pm}$ production by CHORUS~\cite{bib:chorus} is in agreement
with this rate. Using the results of our second fit, we find a rate of (3.2 
$\pm$ 0.6) $\times $ 10$^{-3}$, which is 

\begin{figure}[tbp]                                                             
  \centerline{\psfig{figure=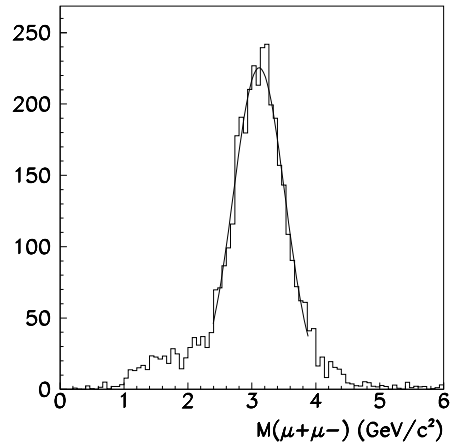,width=6cm}}                          
  \vspace{0.5cm}                                                                
\caption{The two muon invariant mass ($M_{\mu\mu}$) for the                     
$J/\psi$ Monte Carlo. The curve shows a Gaussian fit. }                         
\label{fig:jpsimmumu}                                                           
\end{figure} 

\noindent
consistent with previous results.

Table~\ref{tab:leheventresults} lists the number of events contribution         
of each source in the low-E$_{HAD}$ two muon data sample as determined
by this analysis.   

\subsection{Neutral-Current Analysis \label{sec:jpsimeas}}

Neutral-current $J/\psi $ production produces a clear signature in the 
two muon invariant mass, particularly if $E_{HAD}\leq 3$ GeV\ is imposed to
select diffractively produced events. There is no evidence for a $J/\psi $
signal in Fig. \ref{fig:mmumu}; however, the relatively poor resolution of
the NuTeV detector may be obscuring a contribution from this source. To
assess this possibility, a diffractive $J/\psi $ sample was simulated via
Monte Carlo to obtain the $M_{\mu \mu }$ distribution shown in Fig.~\ref
{fig:jpsimmumu}. A Gaussian fit to this distribution yields a resolution 
$\sigma _{0}$ = 0.40 GeV/c$^{2}$.

A maximum likelihood fit was then performed to determine the amount of 
$J/\psi$ present in the data. The fit function was taken to be 
\begin{equation}
N(M_{\mu \mu })=M_{\mu \mu }^{\alpha }e^{(\beta +\gamma M_{\mu \mu
})}+A\times e^{-\frac{1}{2}(\frac{M_{\mu \mu }-M_{0}}{\sigma _{0}})^{2}},
\end{equation}
where $M_{\mu\mu}$ is the two muon invariant mass. $M_{0}$ and 
$\sigma _{0}$ are the mass and width of the $J/\psi$ as measured by
the Monte Carlo. The first term represents a smooth parameterization of the
background description where $\alpha$ and $\gamma$ determine the shape and $
\beta$ the normalization. The second term is a Gaussian description of the 
$J/\psi$ contribution with mean mass $M_{0}$ and width $\sigma _{0}$
set to the Monte Carlo prediction. The parameter $A$ measures the amount of 
$J/\psi$ in the data.

The results of the fit are shown in Table~\ref{tab:diffjpsi}. A 90$\%$
confidence level (CL) on the $J/\psi$ contribution is set by 
fixing the 
$J/\psi$ amplitude to various increasing levels and fitting for the
background. The likelihood function ({${\cal L}$}(A)) was plotted as a
function of $A$ and the 90$\%$ CL limit set by $\int_{A_{0}}^{A_{CL}}{\cal L}
(A)\ dA/\int_{A_{0}}^{\infty }{\cal L}(A)\ dA=0.90.$ \ The 

\begin{figure}[tbp]                                                             
  \centerline{\psfig{figure=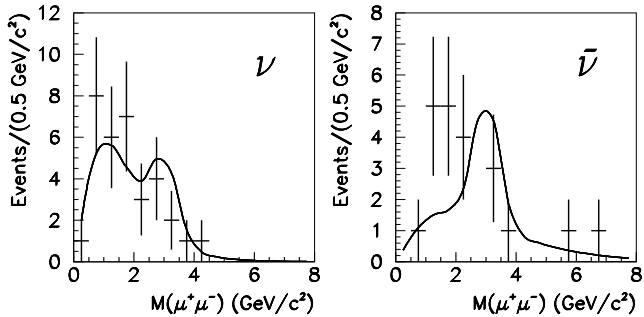,width=\columnwidth}}                 
\caption{90$\%$ confidence level limit on diffractive $J/\psi$                  
production. The curve shows the fit to the background with the 90$\%$ CL        
contribution from $J/\psi$.}                                                    
\label{fig:diffjpsilimit}                                                       
\end{figure}   

\noindent
resulting number
of events is found by integrating the Gaussian with amplitude $A_{CL}$ and
converting to a cross-section by normalizing to the DIS two-muon sample. The
90$\%$ confidence level limits on the diffractive $J/\psi$ cross-section are 
\begin{eqnarray*}
\sigma \left( \nu _{\mu }Fe\rightarrow \nu _{\mu }J/\psi Fe\right) &\leq
&0.21\ \text{fb}/\text{nucleon at }90\%\text{ CL,} 
\end{eqnarray*}
\begin{eqnarray*}
\sigma \left( \bar{\nu}_{\mu }Fe\rightarrow \bar{\nu}_{\mu }J/\psi Fe\right)
&\leq &0.36\ \text{fb}/\text{nucleon at }90\%\text{ CL,}
\end{eqnarray*}
at a mean production energy of $E_\nu$ = 175 GeV for $E_{HAD}\leq 3$
GeV. Figure~\ref{fig:diffjpsilimit} shows the results of this limit
including the $J/\psi$ contribution.

\begin{figure}[tbp]                                                             
  \centerline{\psfig{figure=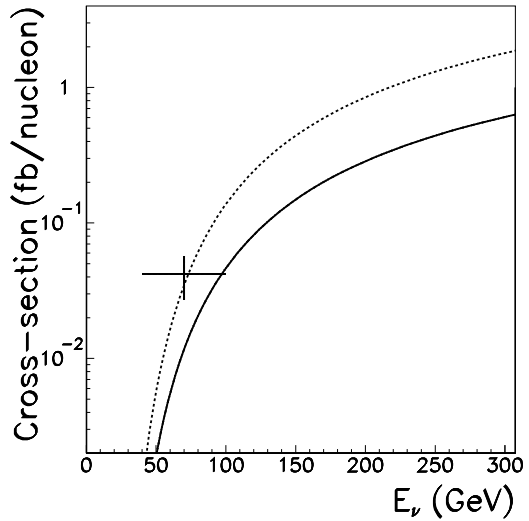,width=7cm}}                      
  \vspace{0.5cm}                                                                
\caption{Energy dependence of the 90$\%$ confidence level limit on              
diffractive $J/\psi$ production normalized to the NuTeV                         
measurement. The solid line is the limit set with E$_{HAD}$ $<$ 3               
GeV and the dashed line with E$_{HAD}$ $<$ 10 GeV. The data point               
shows the CDHS measurement. }                                                   
\label{fig:jpsicompare}                                                         
\end{figure}

Using the VMD model, we can extrapolate the $J/\psi$ cross-section
to lower $E_{\nu }$ to compare to the CDHS measurement of (0.042 $\pm $
0.015) fb/nucleon. Figure~\ref{fig:jpsicompare} shows a 90$\%$ CL limit for
the $J/\psi$ cross-section normalized to the NuTeV measurement and
compares to the CDHS result. The energy dependence of this limit is
dependent upon the model inputs. We show the limit for a 
momentum transfer squared ($t$) distribution of 
$e^{-145t}$~\cite{bib:e632} 
and coherent production only. The interpretation of the 

\begin{minipage}{17.6cm}
 \begin{center}
  \begin{minipage}{12cm}
\begin{table*}[tbp]                                                             
\caption{Limits on diffractive $J/\psi$ production observed by                  
NuTeV in the low-E$_{HAD}$ two-muon sample. }                                   
\label{tab:diffjpsi}                                                            
\begin{tabular}{rcc}                                                            
& $\nu$ mode & $\bar{\nu}$ mode \\                                              
\multicolumn{1}{l}{E$_{HAD}$ $<$ 3 GeV} &  &  \\                                
$\#$ of events (fit) & 3.3 $\pm$ 5.0 & -1.7 $\pm$ 4.0 \\                        
$\#$ of events (90$\%$ CL) & 7.5 & 5.0 \\                                       
Average $E_\nu$ ($J/\psi$) (GeV) & 185.0 & 168.0 \\                             
Cross-Section (90$\%$ CL) ($E_\nu$ = 175 GeV) (fb/nucleon) & 0.19 & 0.32  \\    
Cross-Section (90$\%$ CL) ($E_\nu$ = 70 GeV) (fb/nucleon) & 0.011 & 0.017 \\    
&  &  \\                                                                        
\multicolumn{1}{l}{E$_{HAD}$ $<$ 10 GeV} &  &  \\                               
$\#$ of events (fit) & 7.8 $\pm$ 14.3 & -2.4 $\pm$ 10.5 \\                      
$\#$ of events (90$\%$ CL) & 24.8 & 11.5 \\                                     
Cross-Section (90$\%$ CL) ($E_\nu$ = 175 GeV) (fb/nucleon) &                    
0.63 & 0.73 \\                                                                  
Cross-Section (90$\%$ CL) ($E_\nu$ = 70 GeV) (fb/nucleon) &                     
0.034 & 0.040 \\  
&  &  \\                                                                        
\multicolumn{1}{l}{CDHS result (E$_{HAD}$ $<$ 10 GeV)} &  &  \\                 
Cross-Section (Fit) (($E_\nu$ = 70 GeV) (fb/nucleon) &                          
\multicolumn{2}{c}{$(0.042 \pm 0.015) $}                                        
\end{tabular}                                                                   
\end{table*}                                                                    
  \end{minipage}
 \end{center}
\end{minipage}  

\noindent
CDHS measurement as diffractive production
is contradicted by our limit; however, the CDHS\ $E_{HAD}$ cut of 
$10$ GeV compared
to a mean neutrino energy of 70 GeV could have accepted DIS production of 
$J/\psi$.

The analysis was repeated with a higher E$_{HAD}$ cut (E$_{HAD}$ $<$ 10 GeV) 
which matches the CDHS selection. The results
are also included in Table~\ref{tab:diffjpsi}. The limit as a function of 
$E_\nu$ is shown as the dashed curve in Fig.~\ref{fig:jpsicompare}.
While this limit is near the CDHS measurement, the diffractive 
$J/\psi$ Monte Carlo shows the lower E$_{HAD}$ ($<$3 GeV)
is appropriate for coherent diffractive production. One possible explanation
is that the higher cut allows a contribution from DIS $J/\psi$ production.

\newpage

\begin{figure}[tbp]                                                             
  \centerline{\psfig{figure=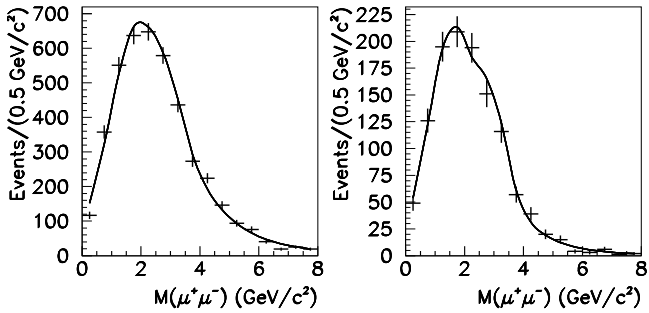,width=\columnwidth}}              
\caption{The two muon invariant mass distribution for high-E$_{HAD}$            
(E$_{HAD}$ $>$ 10 GeV).  The curve shows the fit to the background with         
the 90$\%$ CL contribution from $J/\psi$.}                                      
\label{fig:disjpsilim}                                                          
\end{figure}  

For completeness, the $E_{HAD}$ cut was changed to $E_{HAD}>10$ GeV to allow
a search for inclusive DIS $J/\psi $ production. Figure~\ref{fig:disjpsilim}
shows the resulting $M_{\mu \mu }$ distribution. A fit was performed with
the background modelled by an asymmetric Gaussian function and the data
binned in 0.5 GeV/c$^{2}$ intervals. No statistically significant $J/\psi $
signal was found, and 90$\%$ confidence level limits for the (DIS 
$J/\psi$)/(DIS CC charm) rates were found to be 
\begin{eqnarray*}
\frac{N\left( \nu _{\mu }N\rightarrow \nu _{\mu }J/\psi X\right) }{N\left(
\nu _{\mu }N\rightarrow \mu ^{-}cX\right) } &\leq &0.024\text{ at }90\%\text{
CL,} 
\end{eqnarray*}
\begin{eqnarray*}
\frac{N\left( \bar{\nu}_{\mu }N\rightarrow \bar{\nu}_{\mu }J/\psi X\right) }{
N\left( \bar{\nu}_{\mu }N\rightarrow \mu ^{+}\bar{c}X\right) } &\leq &0.069
\text{ at }90\%\text{ CL,}
\end{eqnarray*}
when averaged over the NuTeV beam spectra. Assuming a similar energy
dependence to DIS CC charm production, this corresponds to cross-section
limits evaluated at $E_{\nu }=125$ GeV, 
\begin{eqnarray*}
\sigma \left( \nu _{\mu }N\rightarrow \nu _{\mu }J/\psi X\right) &\leq &2.2\ 
\text{fb/nucleon at }90\%\text{ CL}, 
\end{eqnarray*}
\begin{eqnarray*}
\sigma \left( \bar{\nu}_{\mu }N\rightarrow \bar{\nu}_{\mu }J/\psi X\right)
&\leq &3.4\text{ fb/nucleon at }90\%\text{ CL},
\end{eqnarray*}
at an average DIS CC charm production energy of $E_\nu$ = 125 GeV.
Figure~\ref{fig:disjpsilim} shows the 90$\%$ CL curve for the inclusive (
E$_{HAD}$ $>$ 10 GeV) DIS two-muon sample.

\section{Conclusions}

We have performed an inclusive analysis of the low-E$_{HAD}$ 
two-muon data sample available from NuTeV. All known Standard Model processes
have been considered, and four significant sources were found to contribute:
DIS, neutrino trident, diffractive $D_S^\pm$, and diffractive 
$D_S^{*\pm}$. We have measured diffractive $D_S^\pm$ and 
$D_S^{*\pm}$ cross-sections to be $1.4\pm 0.4$ fb/nucleon and 
$1.6\pm 0.5$ fb/nucleon, respectively, for $\nu _{\mu }Fe$ scattering at 
$E_{\nu }=130$ GeV, in agreement with previous results assuming a VMD energy
dependence of the cross-section. We observe a statistically weak neutrino
trident signal consistent with Standard Model predictions. Finally, we see
no evidence for either diffractive or DIS production of $J/\psi $ in NC 
$\nu _{\mu }$ and $\bar{\nu}_{\mu }$ scattering on iron.

\acknowledgments
We would like to thank the staffs of the Fermilab Particle Physics and Beams
Divisions for their contributions to the construction and operation of the
NuTeV beamlines. We would also like to thank the staffs of our home
institutions for their help throughout the running and analysis of NuTeV.
This work has been sponsored by the U.S. Department of Energy and the
National Science Foundation.


\begin{references}

\bibitem{bib:asratyan}  A.~E.~Asratyan {\it et al.}, {Z. Phys.} {\bf C 58}, 55 (1993).

\bibitem{bib:chorus}  P.~Annis {\it et al.}, {Phys. Lett.} {\bf B 435}
458 (1998).

\bibitem{bib:trid1}  R.~Belusevic and J.~Smith, {Phys. Rev.} {\bf D 37},
2419 (1987).

\bibitem{bib:trid2}  C.~H.~Llewellyn~Smith, {Phys. Rept.} {\bf 3} 261 (1972).

\bibitem{bib:cdhsjpsi}  H.~Abramowicz {\it et al.}, {Phys. Lett.} {\bf %
109B}, 115 (1982).

\bibitem{bib:labe}  W.~S.~Sakumoto {\it et al.}, {Nucl. Instrum.
Methods.} {\bf A 294}, 179 (1991); B.~J.~King {\it et al.}, {Nucl.
Instrum. Methods.} {\bf A 302} 254 (1991).

\bibitem{bib:ssqt}  R.~H.~Bernstein {\it et al.}, NuTeV Collaboration, 
``Technical Memorandum:  Sign Selected Quadrupole Train,'' FERMILAB-TM-1884 
(1994); J.~Yu {\it et al.}, NuTeV Collaboration, ``Technical Memorandum:  
NuTeV SSQT performance,'' FERMILAB-TM-2040 (1998).

\bibitem{bib:nutevnim}  D.~Harris, J.~Yu {\it et al.}, HEP-EX/9908056,
(submitted to Nucl. Instrum. Methods A).

\bibitem{bib:moriond99}  T.~Adams {\it et al.}, in {\it Proceedings of the
33rd Rencontres de Moriond, QCD and Hadronic Interactions}, (to be published).

\bibitem{bib:charmii}  P.~Vilain {\it et al.}, {CERN-EP/98-128} (1998).

\bibitem{bib:bazarko}  A.~O.~Bazarko {\it et al.}, {Z. Phys.} {\bf C 65}
189 (1995).

\bibitem{bib:rabin}  S.~A.~Rabinowitz {\it et al.}, {Phys. Rev. Lett.} 
{\bf 70}, 134 (1993).

\bibitem{bib:cdhs}  H.~Abramowicz {\it et al.}, {Z. Phys.} {\bf C 15},
19 (1982).

\bibitem{bib:collins-spiller}  P.~Collins and T.~Spiller, {J. Phys.} 
{\bf G 11}, 1289 (1985).

\bibitem{bib:e531}  N.~Ushida {\it et al.}, {Phys. Lett.} {\bf B206},
375 (1988).

\bibitem{bib:bolton-frag}  T.~Bolton, HEP-EX/970814 (1997).

\bibitem{bib:pdg}  Particle Data Group, {Eur. Phys. J.} {\bf C 3} (1998).

\bibitem{bib:sandler}  P.~H.~Sandler, Ph.D. thesis, University of
Wisconsin-Madison, (1992).

\bibitem{bib:intbosIII}  R.~W.~Brown {\it et al.}, {Phys. Rev.} {\bf D6}
, 3273 (1972).

\bibitem{bib:ccfrtrid}  S.~R.~Mishra {\it et al.}, {Phys. Rev. Lett.} 
{\bf 66}, 3117 (1991).

\bibitem{bib:charmiitrid} G.~Geiregat {\it et al.}, {Phys. Lett.} {\bf 
B 245}, 271 (1990).

\bibitem{bib:charmtrid} F.~Bergsma {\it et al.}, {Phys. Lett.} {\bf B
122}, 185 (1983).

\bibitem{bib:fuji1}  K.~Fujikawa, {Ann. Phys.} {\bf 68}, 102 (1971);
K.~Fujikawa, Ann. Phys. {\bf 75}, 491 (1973).

\bibitem{bib:fuji2}  K.~Fujikawa, {Phys. Rev.} {\bf D 8}, 1623 (1973).

\bibitem{bib:icheptrid}  T.~Adams {\it et al.}, in {\it Proceedings of the 
29th International Conference on High Energy Physics}, Vancouver, Canada,
1998, edited by Alan Astbury, David Axen, and Jacob Robinson, p. 631.

\bibitem{bib:e632} S.~Willocq {\it et al.}, {Phys. Rev.} {\bf D47},
2661 (1993).

\bibitem{bib:kopel} B.~Z.~Kopeliovich and P.~Marage, {Int. Jour. Mod.
Phys.} {\bf A 8}, 1513 (1993).

\bibitem{bib:pumplin} J.~Pumplin, {Phys. Rev. Lett.} {\bf 64}, 2751 (1990).

\bibitem{bib:rein} D.~Rein and L.~M.~Sehgal, {Ann. Phys.} {\bf 133}, 79 (1981).

\bibitem{bib:shrock}  R.~E.~Shrock and B.~W.~Lee, {Phys. Rev.} 
{\bf D 13}, 2539 (1976); Erratum-ibid {\bf D 14}, 313 (1976).

\end{references}
\end{document}